\documentclass[5p]{elsarticle}

\usepackage{graphicx}
\usepackage{dcolumn}
\usepackage{bm}
\usepackage{booktabs}
\usepackage{array,multirow}
\usepackage{hyperref}  
\hypersetup{breaklinks = true, colorlinks = true, citecolor = blue, linkcolor = blue, urlcolor = blue}
\usepackage{cleveref}

\newcommand{\be}{\begin{equation}}
\newcommand{\ee}{\end{equation}}
\newcommand{\bea}{\begin{eqnarray}}
\newcommand{\eea}{\end{eqnarray}}
\newcommand{\ed}{$\textit{ed}$}
\usepackage[mathlines]{lineno}

\begin{document}


\begin{frontmatter}

\title{Probing short-range correlations in the deuteron \\ via incoherent diffractive J/$\psi$ production with spectator tagging at the EIC }

\author[BNL]{Zhoudunming~Tu}
\author[BNL]{Alexander~Jentsch}
\author[PDS]{Mark~Baker}
\author[China1,China2]{Liang Zheng}
\author[BNL]{Jeong-Hun Lee}
\author[BNL]{Raju~Venugopalan}
\author[MIT]{Or~Hen}
\author[JLAB]{Douglas~Higinbotham}
\author[BNL]{Elke-Caroline~Aschenauer }
\author[BNL]{Thomas Ullrich}

\address[BNL]{Department of Physics, Brookhaven National Laboratory, Upton, NY 11973, USA}
\address[PDS]{Mark D. Baker Physics and Detector Simulations LLC, Miller Place, NY 11764, USA}
\address[China1]{School of Mathematics and Physics, China University of Geosciences (Wuhan), Wuhan 430074, China}
\address[China2]{Key Laboratory of Quark and Lepton Physics of the Ministry of Education and Institute of Particle Physics, Central China Normal University, Wuhan 430079, China} 
\address[MIT]{Laboratory for Nuclear Science, Massachusetts Institute of Technology, Cambridge, MA 02139, USA}
\address[JLAB]{Electron Ion Collider Center, Thomas Jefferson National Accelerator Facility, Newport News, Virginia 23606, USA}


\begin{abstract}
Understanding the role of Quantum Chromodynamics in generating nuclear forces is important for uncovering the mechanism of short-ranged nuclear interactions and their manifestation in short range correlations (SRC). The future Electron-Ion-Collider (EIC) at Brookhaven National Laboratory in the US will provide an unprecedented opportunity to systematically investigate the underlying physics of SRC for energies and kinematic regions that are otherwise impossible to reach. We study SRCs in electron-deuteron scattering events using the Monte Carlo event generator BeAGLE. Specifically, we investigate the sensitivity of observables  to high internal nucleon momentum in incoherent diffractive $J/\psi$ vector meson production. In a plane wave impulse approximation, the initial state deuteron wavefunction can be accessed directly from the four-momentum of the spectator nucleon. We use  realistic physics simulations and far-forward detector simulations of the EIC to fully reveal the physics potential of this exclusive process. In particular, we provide the luminosity and detector requirements necessary to study SRCs in the deuteron at an EIC. 
\end{abstract}


\end{frontmatter}

\section{\label{sec:intro} Introduction}
Lepton deep inelastic scattering measurements conducted at accelerator facilities around the world\cite{Aubert:1983xm,Ashman:1988bf,Gomez:1993ri,Arneodo:1988aa,Arneodo:1989sy,Allasia:1990nt,Seely:2009gt,Schmookler:2019nvf} uncovered that the bound nucleons in nuclei are significantly modified in terms of their parton distribution functions (PDFs) in the valence quark region (Bjorken-$x$ within $0.3<x<0.7$). 
This is known as the ``EMC effect".  
The microscopic origins of the EMC effect are still not fully understood nearly four decades after its original discovery~\cite{Norton03,Higinbotham:2013hta,Malace:2014uea,Hen:2016kwk}.

More recently, quasielastic proton and lepton scattering measurements uncovered another remarkable phenomenon, that of short range nucleon-nucleon correlations (SRCs)~\cite{Hen:2016kwk,ciofi15}. 
SRCs are pairs of strongly interacting nucleons at close proximity. They dominate the high-momentum distribution of the many-body nuclear wave function and exhibit universal behavior in nuclei from Deuterium to Lead~\cite{tang03,piasetzky06,subedi08,korover14,Hen:2014nza,duer18,Duer:2018sxh,schmidt20}. 

Experimental data from JLab suggest a strong link between SRCs and the EMC effect~\cite{Schmookler:2019nvf,Hen:2016kwk,weinstein11,Hen12,Hen:2013oha}. Specifically, they suggest that the underlying mechanism of nucleon modifications could be caused by short-range correlated nucleon pairs with high internal nucleon momentum, for instance, a quasi-deuteron inside the nucleus. However there are alternative phenomenological models that can explain the EMC effect without involving SRCs; see Ref.~\cite{Hen:2016kwk} for a recent review. 

The difficulty in drawing a definitive conclusion based on available experimental data is primarily due to the complexity of the nuclear environment. Given the differing structure and reaction dynamics of different nuclei, the nuclear mass ($A$) dependence could in principle be attributed to other underlying physical mechanisms. Nuclear effects that are driven by SRCs should be similar in light nuclei at extreme high internal nucleon momentum and in medium and heavy nuclei in a similar kinematic range. Therefore the observation of universal properties across a wide range of nuclei would suggest that the effect may be independent of the specifics of nuclear structure and reactions. A confirmation of such universal behavior would then provide a definitive explanation for the EMC puzzle. It may also provide insight into similarly universal dynamics, independent of microscopic details, in physical systems across varying energy scales. 

Besides the modifications in the valence quark region in the bounded nucleon, there are a number of  other outstanding questions:
\begin{itemize}
\item What role do gluons play in the short-range correlations of  nucleon pairs?
\item Are gluon modifications linked to the SRC, similar to that for  valence quarks?
\item What is the relation of SRCs to gluon shadowing ?
Can this be related to the phenomenon of gluon saturation? 
\item What are the spatial and momentum distributions of partons in such high nucleon momentum configurations ?
\end{itemize}
With regard to the last item, nucleon-nucleon elastic scattering experiments at high momentum transfer showed that the energy dependence of such reactions is quite sensitive to differing models of the internal spatial and momentum distributions of partons~\cite{Sterman:2010jv}. 

In recent years, among other probes of the gluon distributions at low-$x$, the LHC experiments have observed significant gluon shadowing effects in Pb nuclei via coherent diffractive $J/\psi$ photoproduction in ultra-peripheral collisions (UPC)~\cite{Khachatryan:2016qhq,Abelev:2012ba}. The UPC results strongly suggest that the gluon distributions, $xG(x,Q^{2})$, are significantly suppressed compared to the proton in $x$ regions~\cite{Khachatryan:2016qhq,Strikman:2018mbu} where gluons are dominant. 

However in UPC events in heavy ion collisions, without the additional lever arm of the momentum transfer, $Q^{2}$, the photoproduction of the  $J/\psi$ can only provide limited information since the only hard scale in the process is set by the mass of the vector meson. Moreover,  in the LHC experiments,   nucleon configurations are not easily accessible using  final state particle spectra, unlike what can be achieved at fixed target experiments~\cite{Egiyan:2007qj}. Therefore in order to answer the aforementioned outstanding questions,  diffractive vector meson production (for example, the $J/\psi$) in a simple nucleus, over a wide range of kinematic phase space and well-controlled initial nucleon configurations, can be an extremely effective experimental measurement. 

\begin{figure}[t]
\includegraphics[width=\linewidth]{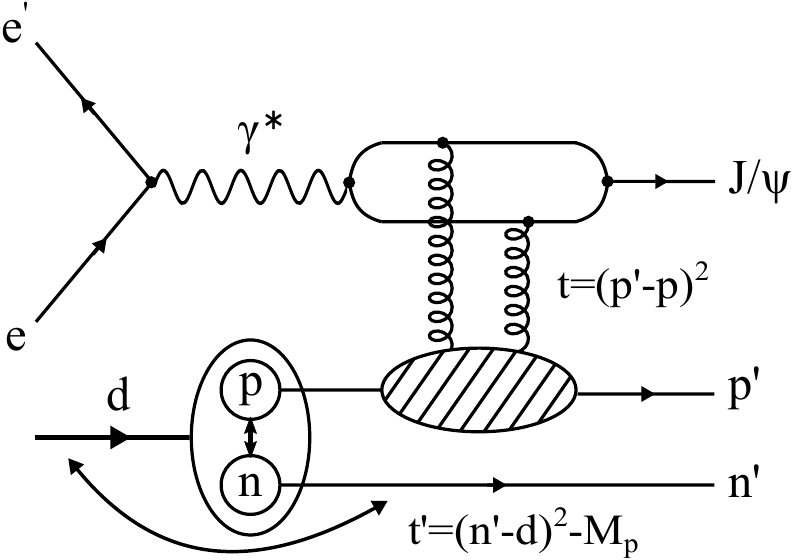}
  \caption{ \label{fig:figure_1} Diagram of incoherent diffractive $J/\psi$ productions in electron-deuteron scattering }
\end{figure}

In January 2020, the US Department of Energy announced the Critical-Decision 0 mission need for a high energy and high luminosity Electron-Ion Collider (EIC). At the EIC, we propose to look at the process of incoherent diffractive $J/\psi$ production in electron-deuteron (\ed) collisions by fully reconstructing the final state shown in Fig.~\ref{fig:figure_1}. In this process, the primary interaction is between the virtual photon and one of the bound nucleons, either the proton or the neutron, with only a diffractive $J/\psi$ meson produced in addition to  the breakup nucleons in the forward region.  

In the  plane wave impulse approximation (PWIA), the spectator nucleon carries the initial nucleon momentum information of the bound nucleon pairs without any distortion from the interaction. This is key to determining the initial state configuration of the deuteron on an event-by-event basis. Diffractive $J/\psi$ production can be investigated in different configurations, and because the properties of the $J/\psi$ are relatively well understood, can provide a clear theoretical understanding of gluon mediated processes. 

In an analogous process, introduced in Ref.~\cite{Miller:2015tjf}, the primary interaction is via  pomeron exchange with both bound nucleons simultaneously with the proton and the neutron remaining intact in the final state with large relative momentum transfer between them.  Thus to preserve color neutrality, and preserve the proton-neutron final state, an additional color octet gluon exchange must take place between the  two nucleons. 

In this paper, only simulations based on the PWIA picture are presented, where the proton-neutron SRC is mediated by a color singlet exchange. However our results will also provide considerable insight in general into uncovering the physics underlying gluon mediated (color octet) short range correlations in the deuteron.

The outline of this paper is as follows. In Sec.~\ref{sec:kinematics}, we begin by defining the kinematic variables that are relevant for the process of interest. In Sec.~\ref{sec:deuteron}, the deuteron wavefunction in a light front (LF) formulation of quantum mechanics is introduced briefly, followed by its implementation in the Benchmark eA Generator for Leptoproduction (BeAGLE) event generator in Sec.~\ref{sec:model}. In Sec.~\ref{sec:design}, a brief introduction is given of the Interaction Region (IR) design (based on the EIC pre-CDR design) and of the proposed forward particle detectors. In Sec.~\ref{sec:result}, physics simulations and detector simulations are provided. Sec.~\ref{sec:conclusion} summarizes the results of the paper.

\section{\label{sec:kinematics} Kinematics}

The kinematics of the reaction $e+d\rightarrow e'+J/\psi+p'+n'$ are defined by the following variables: the negative four-momentum transfer squared at the electron vertex, $Q^{2}=-q^{2}=-(e-e')^{2}$; the momentum transfer squared at the leading nucleon vertex, $t=(p'-p)^{2}$, and the inelasticity $y=(p\cdot q)/(p\cdot e)$; the momentum transfer squared between the spectator and the incoming deuteron minus the leading nucleon mass, $t'=(n'-d)^{2}-M_{p}$. The four-momentum $e$, $e'$, $p$, $p'$, $n$, $n'$, $d$, and $q$ respectively refer to the incident and scattered electron, the incoming and outgoing leading nucleon, incoming and outgoing spectator nucleon, the incoming deuteron, and the exchanged photon. Note that the leading nucleon can be either the proton or the neutron. The center-of-mass energy $W_{\gamma N}$ between the virtual photon and the leading nucleon is given by, $W^{2}_{\gamma N}=(p+q)^{2}$, where the center-of-mass energy $W_{\gamma d}$ between the virtual photon and the deuteron is given by, $W^{2}_{\gamma d}=(d+q)^{2}$. 

The kinematic variable $t$ is defined between the four-momentum of the incoming and outgoing leading nucleon, while the incoming nucleon momentum inside of the deuteron is not known directly due to the internal nucleon momentum distribution. This is different from the process of electron-proton ($ep$) scattering where the incoming proton has the beam momentum. In an $ep$ collider experiment, the paradigmatic example thus far being the H1 and ZEUS experiments at HERA, the $t$ variable can in principle be reconstructed using the following methods:
\begin{enumerate}
\item $t=(p'-p)^{2}$, if the outgoing proton can be reconstructed using the far-forward proton spectrometer. 
\item $t=(e-e'-V)^{2}$, identical to the previous method, but now  reconstructed from the scattered electron and vector meson ($V$) four-momenta.
\item $t \approx (p_{\rm{_{T,V}}} + p_{\rm{_{T,e'}}})^{2}$, where $p_{\rm{_{T,V}}}$ and $p_{\rm{_{T,e'}}}$ are the transverse momenta of the vector meson and the scattered electron respectively.
\end{enumerate} 

For the $ed$ scattering case, because of the Fermi momentum of the nucleons, method 1) is not applicable even with full reconstruction of the outgoing leading nucleon. Method 2) is kinematically identical to method 1), where the resolution of reconstructing the variable $t$ depends on the detector resolution of reconstructing the four-momentum of the vector-meson and the scattered electron. Generally, this method has poor resolution at very low $t$ because one has to subtract large numbers, for instance, the energy of the electron. Method 3) is similar to method 2), and is an approximate method of only using the transverse component of the momentum to reconstruct the $t$ variable. The accuracy of method 3) relative to method 2) depends on both the detector resolution of final state particles and the $Q^{2}$ of the event. 

In this analysis, we propose to reconstruct the $t$ variable using a novel spectator tagging method, where the initial incoming leading nucleon momentum can be obtained via the spectator four-momentum. This  assumes that the proton and neutron  in their own rest frame are back-to-back before the interaction. One then reconstructs the variable $t$ as 
\be
t = (p'-(-n))^{2}.
\ee
\noindent This reconstruction method 
uses only the forward going nucleons, which are decoupled from the scattered electron and the vector meson. The resolution of the variable $t$ depends on the forward detector resolutions and effects related to the colliding beams (e.g. beam angular divergence). Comparisons among different methods are summarized in Sec.~\ref{sec:appendix}.

\section{\label{sec:deuteron} Deuteron Light Front Wavefunctions}
In this section, we will employ a deuteron spectral function analysis that is identical to that developed to describe the deuteron bound state in the light front (LF) framework, details of which are given in ~\cite{Strikman:2017koc}. 

The non-relativistic nucleon momentum distribution $n(k)$ is taken from Ref.~\cite{CiofidegliAtti:1995qe}. In LF kinematics, 
\ed\ scattering is defined by variables~\cite{Frankfurt81,Frankfurt88,piasetzky06}: 
\begin{itemize}
 \item $p^{+}=(E+p_{z})/\sqrt{2}$, 
  \item $p^{-}=(E-p_{z})/\sqrt{2}$, 
  \item $p_{T}=\sqrt{p^{2}_{x}+p^{2}_{y}}$.
\end{itemize}
In addition, the light cone momentum fraction of the leading nucleon is given by  $\alpha_{p} = 2p^{+}_{p}/p^{+}_{d}$, 
and that of the spectator nucleon is $\alpha_{n} = 2-\alpha_{p}$, where we assume the convention to be $p^{+}_{d}>0$. 

For the two-body deuteron bound state, the LF wave function can be expressed in a 3-dimensional rotationally invariant form. This  allows one to approximate the LF wave function in terms of the 3-dimensional non-relativistic wave function and the analytic properties of the LF  nucleon momentum distribution.

Starting from a $pn$ center-of-mass frame, the nucleon momenta of the  proton and the neutron are back-to-back, denoted by the 3-momentum $k$. The transformation from the $pn$ center-of-mass frame to the 
deuteron rest frame in terms of the light cone momentum fraction and transverse momentum $p_{T}$ of each nucleon is given by

\be
k_{x} = p_{x},k_{y} = p_{y},k_{z} = (\alpha_{n}-1)\sqrt{\frac{(p^{T}_{n})^{2}+M^{2}_{N}}{\alpha_{n}(2-\alpha_{n})}},\\
\ee
\be
E_{N} = \sqrt{k^{2}+M^{2}_{N}}.
\ee 
\noindent where the $M_{N}$ is the average rest mass of the proton and the neutron. Using the LF variables we introduced, the energy $E_{p}$ and the longitudinal momentum $p_{z,p}$ of the leading nucleon, and the the energy $E_{n}$ and longitudinal momentum  $p_{z,n}$  of the spectator nucleon,  can be expressed respectively as 
\be
\label{eq:7}
E_{p} = \frac{(\alpha_{p}M_{d})}{4} + \frac{(p^{2}_{x}+p^{2}_{y}+M^{2}_{p})}{(\alpha_{p}M_{d})};
\ee
\be
\label{eq:8}
p_{z,p} = -\frac{(\alpha_{p}M_{d})}{4} + \frac{(p^{2}_{x}+p^{2}_{y}+M^{2}_{p})}{(\alpha_{p}M_{d})};
\ee
\be
\label{eq:9}
E_{n} = \frac{(\alpha_{n}M_{d})}{4} + \frac{(p^{2}_{x}+p^{2}_{y}+M^{2}_{n})}{(\alpha_{n}M_{d})};
\ee
\be
\label{eq:10}
p_{z,n} = -\frac{(\alpha_{n}M_{d})}{4} + \frac{(p^{2}_{x}+p^{2}_{y}+M^{2}_{n})}{(\alpha_{n}M_{d})}.
\ee

Note that while the light front formalism developed in \cite{Strikman:2017koc} was applied previously to the \ed\ inclusive deep inelastic scattering (DIS) process with spectator tagging, it also applies to the exclusive process considered here, in particular with regard to the description of the deuteron wave function.

\section{\label{sec:model} BeAGLE simulations}
BeAGLE is a general purpose lepton-nucleus ($eA$) event generator;   a detailed description can be found in Ref.~\cite{Beagle}. We will briefly describe below the implementation of the physics of the deuteron in this event generator. The code version (git tag) used in this paper is BeAGLE 1.0.
 
The basic structure of the program is as follows:
\begin{itemize}
\item PYTHIA 6.4~\cite{Sjostrand:2006za} provides the elementary interaction between the virtual photon and the nucleon
\item DPMJET 3.0~\cite{Roesler:2000he} handles the nucleon configurations and geometry for a nucleus with mass $A$ based on a  Glauber model~\cite{Miller:2007ri};
\item FLUKA~\cite{Bohlen:2014buj,Ferrari:2005zk} handles the nuclear remnant and nuclear breakup. 
\end{itemize}

For the process of interest $e+d\rightarrow e'+J/\psi+p'+n'$, the program is much simpler than for nuclei heavier than a deuteron, with DPMJET and FLUKA playing no role. This is because no remnant nucleus is produced following the primary interaction of elastic $J/\psi$ production off either the proton or the neutron with equal probability. 

The electron-neutron scattering is handled the same way as $ep$ in PYTHIA-6, while  adjusted for isospin effects of the neutron relative to the proton. For technical reasons, the nuclear PDF (nPDF) of the deuteron in BeAGLE is currently based on the nPDF of alpha particles ($A=4$). The spectator in this process is consistent with the PWIA, where the four-momentum of the initial state nucleon can be accessed by the final state spectator using Eqs.~\ref{eq:7} - \ref{eq:10}. 

The observables sensitive to the gluon distributions and the SRC studied in this paper are, 
\begin{itemize}
\item $p_{m}$, the 3-momentum of the spectator nucleon;
\item $p_{T}$, the transverse momentum of the spectator nucleon;
\item $\alpha_{p}$ and $\alpha_{n}$, the light cone momentum fraction of the leading and spectator nucleon;
\item $\theta_{\rm{rq}}$, the angle between the virtual photon and the spectator nucleon in the deuteron rest frame.
\item $t$, momentum transfer between the incoming and outgoing leading nucleon,  reconstructed using the novel method we outline that is based on the double tagging of the final state;
\end{itemize}

The center-of-mass energy configuration used in this study is  18 $\mathrm{GeV}$ electrons scattering off 110 $\mathrm{GeV}$ deuterons, where 110 $\mathrm{GeV}$ is the per nucleon energy.  The highest per nucleon energy can be 135 $\mathrm{GeV}$ for deuteron beams at the EIC. The kinematic phase space of this study is based on $1<Q^{2}<10~\rm{GeV^{2}}$ and $0.01<y<0.95$. Note that the physics of the spectator and the deuteron breakup is assumed to be independent of the event kinematics at the lepton vertex.

In addition, BeAGLE simulation can generate events with nucleon dissociation, where the leading nucleon breaks up after the primary interaction. Using both the main detector and forward detectors, 98\% of the dissociative events can be experimentally rejected. We do not discuss this process in the present study.

Due to the design of the BeAGLE event generator, where the hard interaction involves only one nucleon, the coherent diffractive $J/\psi$ production off of the deuteron is currently unavailable. However, the process is similar to the elastic $J/\psi$ production off protons in terms of the kinematic variables and acceptances. 

\section{\label{sec:design} The Experimental Setup at Forward Rapidities}
The current IR and forward region design are based on the pre-Conceptual Design Report (pre-CDR) given by Brookhaven National Lab, with additional, more recent improvements not included in the pre-CDR; see Ref.~\cite{preCDR} for reference.  The detectors are tentative concepts for measuring forward-going particles that are outside the main detector acceptance ($\Theta < $ 35 mrad). Some general considerations used to establish baseline particle acceptance and detector resolutions via full simulations in Geant4~\cite{GEANT4} are presented here.

\begin{figure}[]
\includegraphics[width=\linewidth]{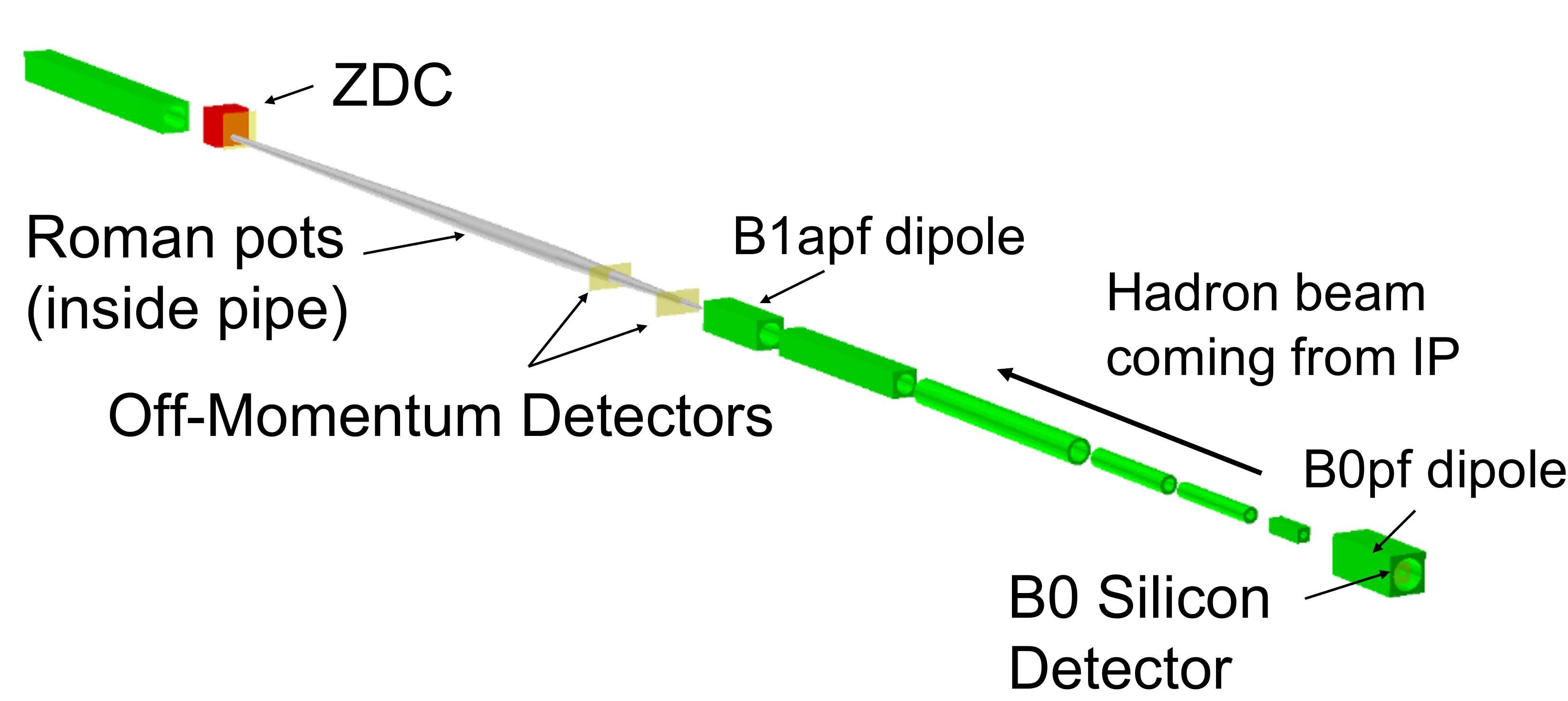}
  \caption{ \label{fig:figure_IR} The layout of the EIC far-forward region depicting the detectors for  proton and neutron detection.}
\end{figure}

In Fig.~\ref{fig:figure_IR}, the geometric layout of the far-forward region of the IR of the hadron-going direction is shown, together with a tentative conceptual design of far-forward particle detectors. The green rectangular boxes denote the dipole magnets; the green cylindrical boxes denote the quadrupole focusing magnets; the gray cylindrical tube is a simple representation of the beam pipe in the drift region where many of the far-forward protons and neutrons are detected. A detailed engineering design of the beam pipe is currently in progress.

\begin{table}[thb]
\fontsize{9}{13}\selectfont
\begin{center}
\begin{tabular}{lccccc}
\hline
  \textbf{Detectors} & Location (x,z) & $\theta$ Acc.  & $p_{T}$ res.   \\
   & [m] &  [mrad] &  [MeV/c]  \\
\hline
 B0 tracker & (0.19, 5.4) & [5.5, 25.0]  & 15-30  \\
 Off-momentum & (0.75, 22.5)  & [0.0, 5.0]   & 20-30 \\
 Roman Pots & (0.85, 26.0) & [0.0, 5.0]  & 20-30  \\
 ZDC &  (0.96, 37.5) & [0.0, 4.5]  & 30-40 \\
\hline
 \end{tabular}
 \end{center}
  \caption{\label{tab:table1} Summary of the physical location, maximum polar angular acceptance, and total transverse momentum resolution smearing for the four far-forward detectors.}
 \end{table}
 
Four different forward detectors are considered in the current study, the red arrows indicate their relative locations in the forward region. They are the $B0$ silicon tracker, the off-momentum Detectors (OMD), the Roman Pots (RP), and the Zero-Degree Calorimeter (ZDC). The location, size, resolution of transverse and total momentum, energy, and scattering angle of these four detectors used in the simmulation are summarized in Table.~\ref{tab:table1}. The numbers in the table reflect maximum acceptances and average resolutions, and do not reflect asymmetries in azimuthal acceptance driven by the size of the aperture, rotation of the individual lattice elements, and optical properties of the lattice. The coordinate system is defined such that the beam is along the z-axis, the x-axis determines the position along the floor transverse to the beam, and the y-axis is the elevation. All detector and beam-lattice components are at the same elevation (i.e. same y-coordinate).

Besides the intrinsic detector resolutions, there are a few beam related effects that induce smearing 
in the reconstructed variables, e.g., the angular divergence of the beam, crab cavity rotation, and the beam momentum spread, as summarized in Table \ref{tab:table2}.

Implementation of detector and beam related effects are included in the following way. For effects related to the proton silicon detectors, they are included in the full-reconstruction. For example, the finite pixel size on the silicon is included by using Gaussian smearing of the location of the hit on the sensor proportional to the desired pixel size. These smeared hits are then used with Kalman filter~\cite{KalmanFilter} reconstruction for the B0 sensors, and a transfer matrix for the OMD and RP. For the OMD and RP, 500 x 500 $\mu m^{2}$ pixels were used in the simulations, while the B0 sensors were simulated with 50 x 50 $\mu m^{2}$ pixels.

As of yet, no full Geant4 level ZDC simulation is included, only realistic acceptance from the lattice layout is determined, while effects such as energy and angular resolution are included by using pure Gaussian smearing. The energy resolution for this study was chosen to be $\frac{\sigma_{E}}{E} = \frac{50\%}{\sqrt{E}} + 5\%$, and the angular resolution was chosen to be $\sigma_{\theta} = \frac{3 \rm{mrad}}{\sqrt{E}}$.

\begin{table}[thb]
\fontsize{8}{13}\selectfont
\begin{center}
\begin{tabular}{lc}
\hline
\noalign{\vskip 1mm}
  \textbf{Beam-related effects} & What is the smearing effect?  \\
  \noalign{\vskip 1mm}
\hline
 Angular Divergence & Hadron beam $\Delta\theta_{h,v}\sim$ 65, 229 $\mu\rm{rad}$     \\
 Crab Cavity & Vtx. $\Delta (x,y) \sim$ (1.25 mm, 0.2 mm) \\
 Beam Energy Spread & Deuteron energy $\sim 10^{-4}$ GeV/$c$     \\
\hline
 \end{tabular}
 \end{center}
  \caption{\label{tab:table2} Summary of the beam-related effects that result in momentum and energy smearing. The various effects are listed from top-to-bottom with the most dominant smearing effect listed first. $\Delta\theta_{h,v}$ is the initial uncertainty of the incoming hadron beam polar angle with respect to the longitudinal axis, where subscript $h$ and $v$ represent horizontal and vertical direction in the transverse plane, respectively.}
 \end{table}
 
For the beam related effects, smearing of the three-momentum components of the nominal deuteron beam are carried out using Gaussian smearing with a width proportional to the values of angular divergence and beam energy spread. The modified deuteron beam four-vector is used to calculate a Lorentz boost vector. The final state protons and neutrons are then boosted from the lab-frame to the deuteron rest frame using the original, un-smeared deuteron boost vector, and then boosted back to the lab-frame using the smeared boost.

\section{\label{sec:result} Results}

In BeAGLE simulations of incoherent diffractive $J/\psi$ meson production in \ed\ scattering, both cases where the spectator nucleon can be either a proton or a neutron are considered. In the simulations, the two cases are treated identically at the generator level, while in the reconstruction of the final state particles in the detector simulations, the spectator proton or neutron would experience different acceptances and detector smearing based on the parameters defined in Table.~\ref{tab:table1} and Table.~\ref{tab:table2}.

\begin{figure}[thb]
\includegraphics[width=1.65in]{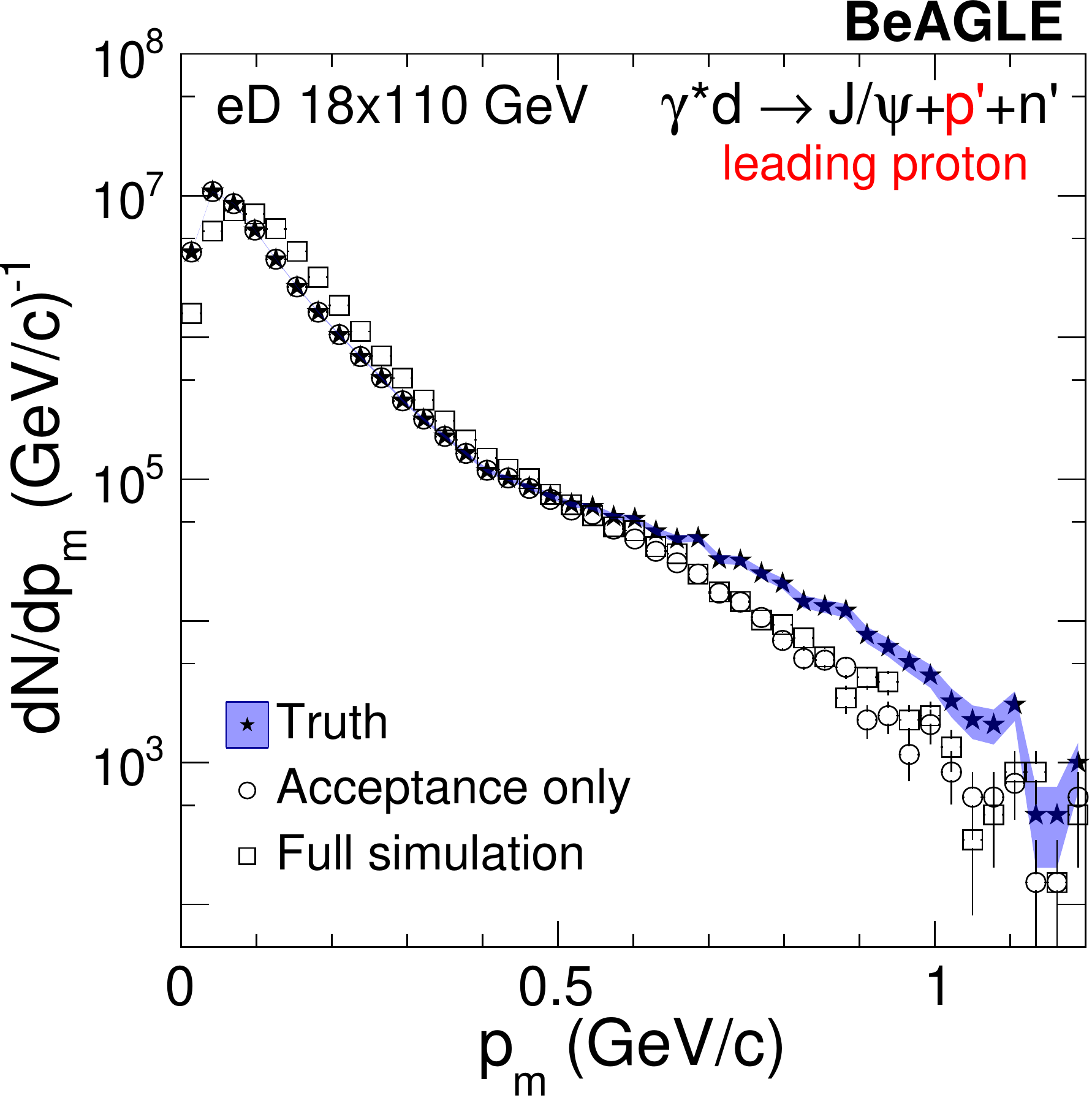}
\hspace{0.02in}
\includegraphics[width=1.65in]{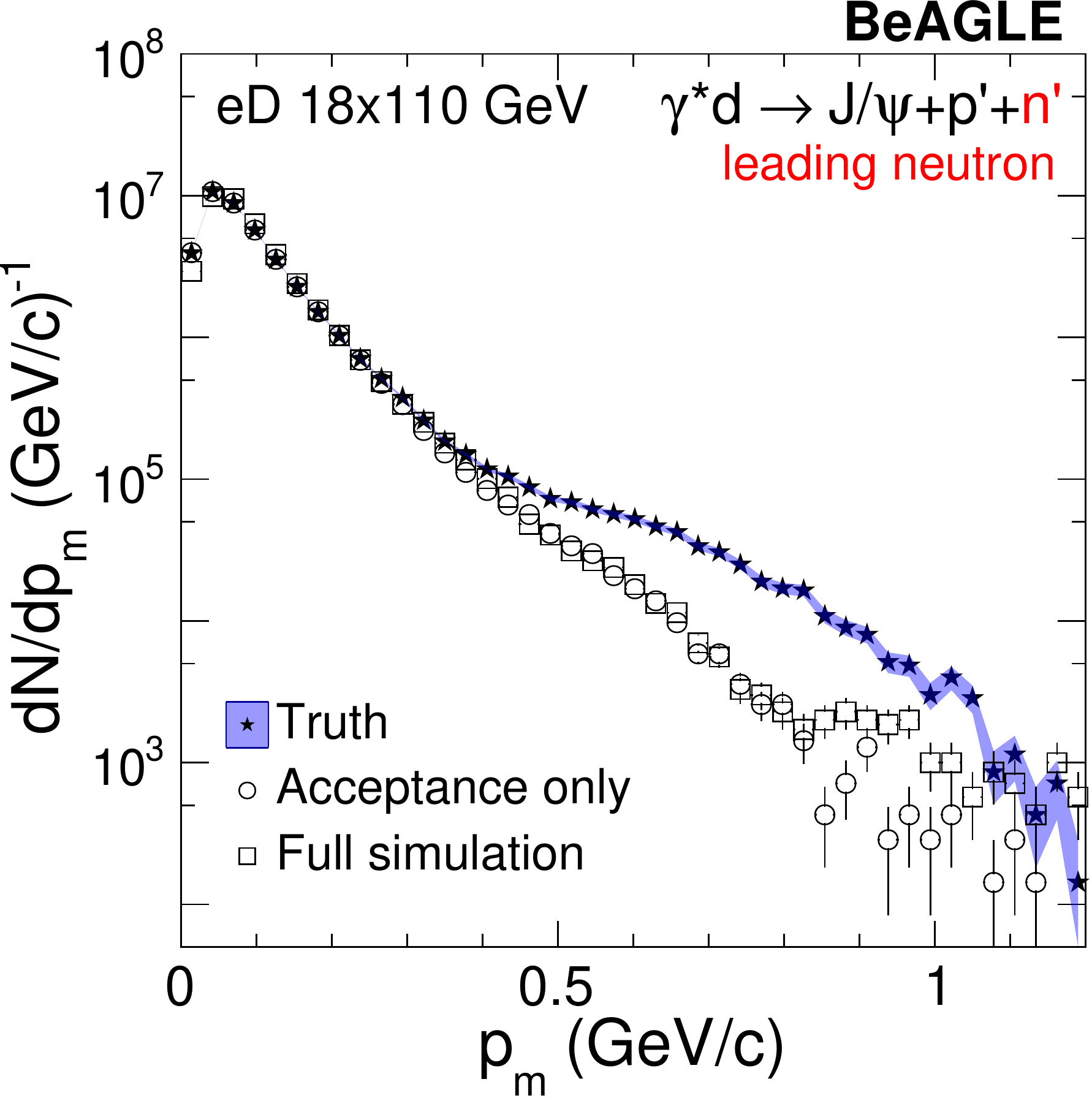}
  \caption{ \label{fig:figure_2} Distribution of the three-momentum of the spectator nucleon  in events associated with incoherent diffractive $J/\psi$ vector meson production in \ed\ collisions are shown for the BeAGLE event generator. The left panel is for the neutron spectator case, where the right panel is for the proton spectators. The simulations at the generator level, with acceptances effects only, and for the full simulations, are shown with solid, open circles, and open squared markers, respectively.}
\end{figure}

In Fig.~\ref{fig:figure_2}, the three-momentum distributions of the spectator, $p_{\rm{m}}$, associated with incoherent diffractive $J/\psi$ production in \ed\ collisions, are shown for neutron  (left) and proton  (right) spectator, respectively. In each panel, the truth level simulation from BeAGLE is shown by solid star markers, where the open circles represent the results after the realistic simulation of the detector acceptance and forward instrumentation. The results of the full simulations (open square markers,)  include acceptances, smearing effects coming from intrinsic detector resolutions, and beam-related effects. The $p_{\rm{m}}$ distribution reflects the internal nucleon momentum at the initial state of the deuteron wave function. For momentum ranges from 0--300 $\rm{MeV/c}$, this region is usually regarded as the mean-field region, while after $p_{\rm{m}}>300~\rm{MeV/c}$, the high momentum tail is regarded as the SRC region.   

The deuteron internal nucleon momentum distributions in the BeAGLE simulation are based on extrapolations of low energy inclusive scattering data~\cite{CiofidegliAtti:1995qe}, as introduced earlier in Sec.~\ref{sec:deuteron}. At the EIC, this measurement of exclusive $J/\psi$ production, together with inclusive \ed\ scattering, will be able to address the following two questions:
\begin{itemize}
\item What is the effective high resolution nucleon-nucleon (NN) potential at extremely high momenta,  $p_{m}\approx1~\rm{GeV/c}$ ?
\item What is the NN potential for events that are only associated with an exclusive $J/\psi$ vector meson? Is the NN potential different when an exclusive channel is probed directly?
\end{itemize}

For the detector and beam-related effect simulations, one sees that the measurements at low momentum would have a larger impact from detector resolutions but with almost 100\% acceptance, however for the high momentum range, the impact is found to be opposite. 

Note even that at the generator level, proton and neutron spectator cases are identical, reflecting the assumptions on the deuteron wave function. However after acceptance effects and detector smearing are applied in the reconstruction, the resulting distributions are different. In the neutron spectator case, most of neutrons reconstructed by the ZDC are within a $\pm 4-6~\rm{mrad}$ cone varying with the azimuthal angle. The non-uniformity of the azimuthal acceptance is due to the aperture of  magnets and the other forward instrumentation. 
The neutron spectator acceptance is almost 100\%  up to 600 $\rm{MeV/c}$, while about 80\% for $p_{m}\approx1~\rm{GeV/c}$. The momentum smearing effect is noticeable for momenta up to 300 $\rm{MeV/c}$. For a nominal beam momentum particle, e.g., $110~\rm{GeV/c}$, the resolution is typically 5\%, dominated by the constant energy resolution term of the ZDC. 

For the proton spectator case on the other hand, the $p_{\rm{m}}$ distributions are found to be different from the neutron. Since the proton has better overall resolution, the $p_{\rm{m}}$ distribution at low-momentum exhibits less bin migration in the tagged proton case, and a better acceptance for high nucleon momenta. Most of the proton spectators end up within the acceptance of the OMD instead of the RP due to the protons having less magnetic rigidity ($\sim 50\%$) compared to the deuteron beam. 

\begin{figure}[thb]
\includegraphics[width=0.49\linewidth]{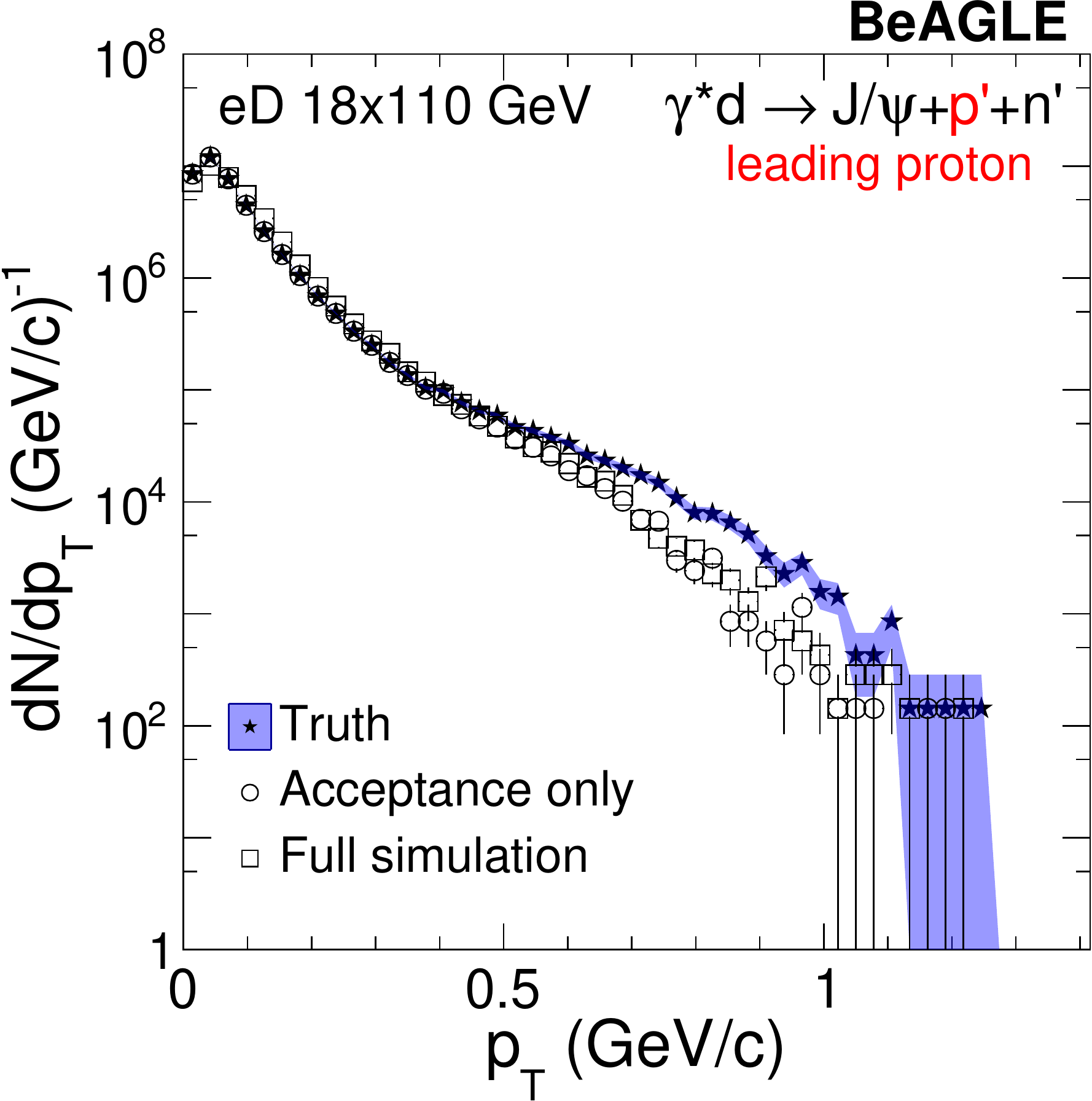}
\includegraphics[width=0.49\linewidth]{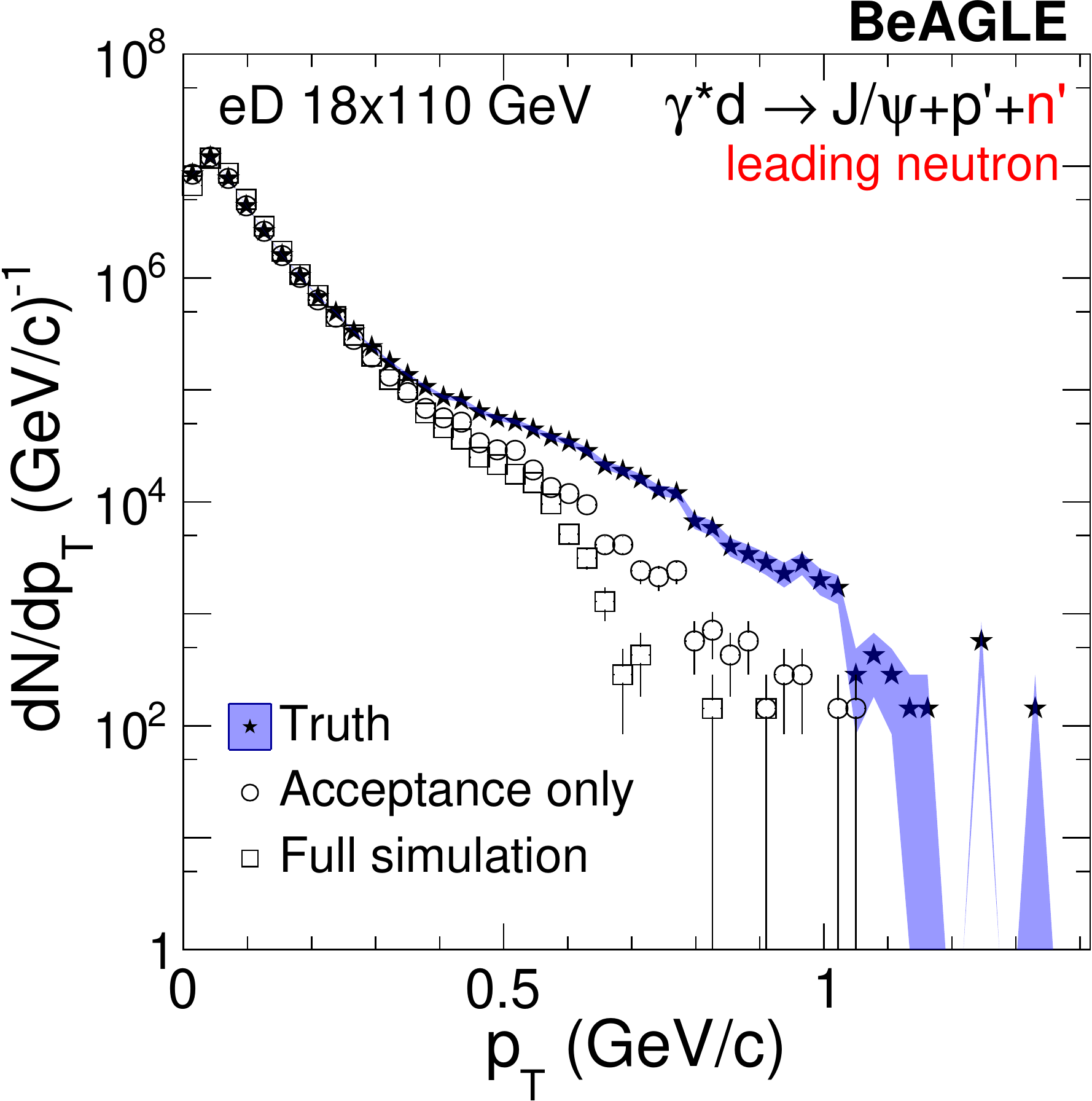}
\includegraphics[width=0.4955\linewidth]{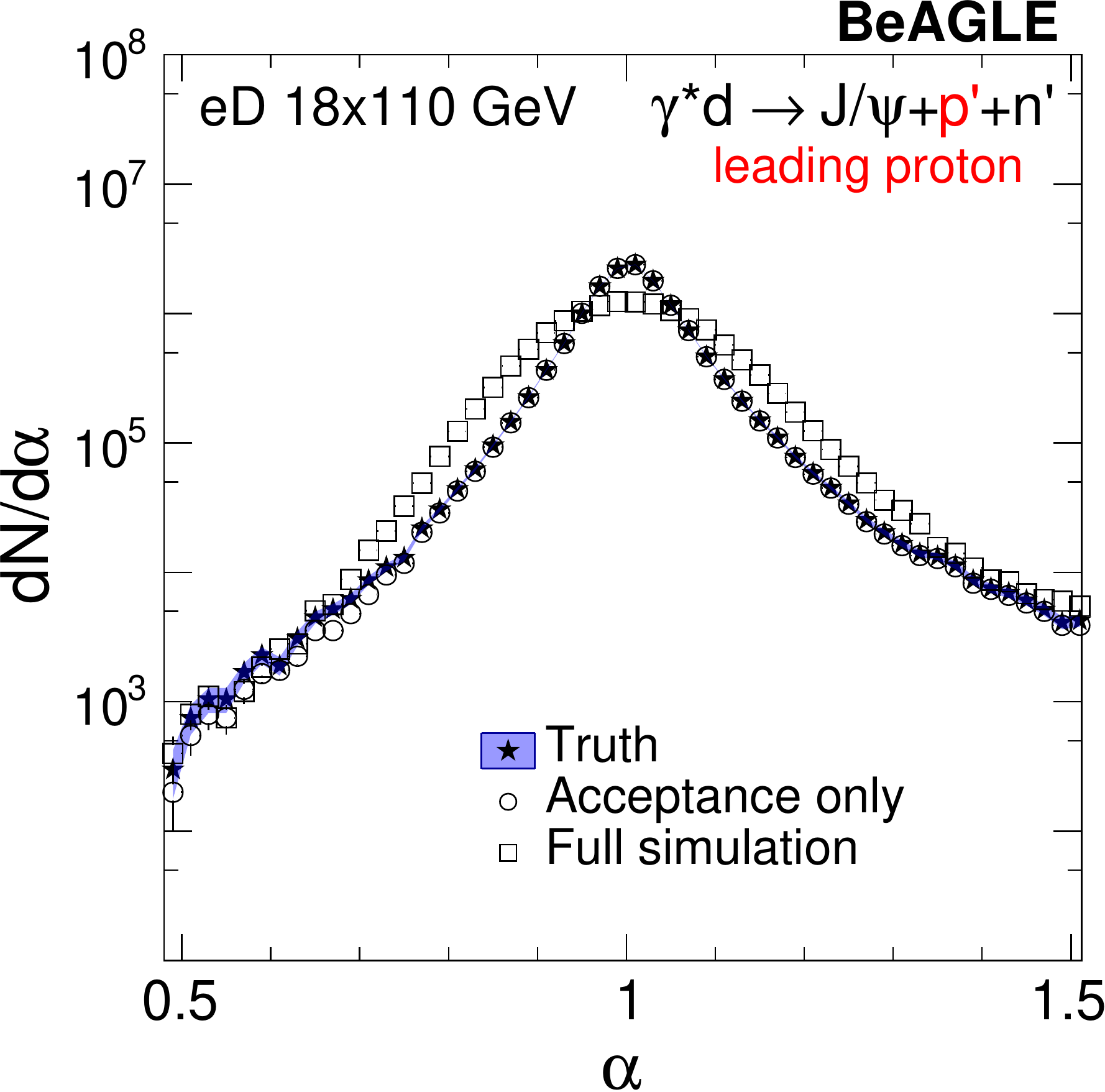}
\includegraphics[width=0.4955\linewidth]{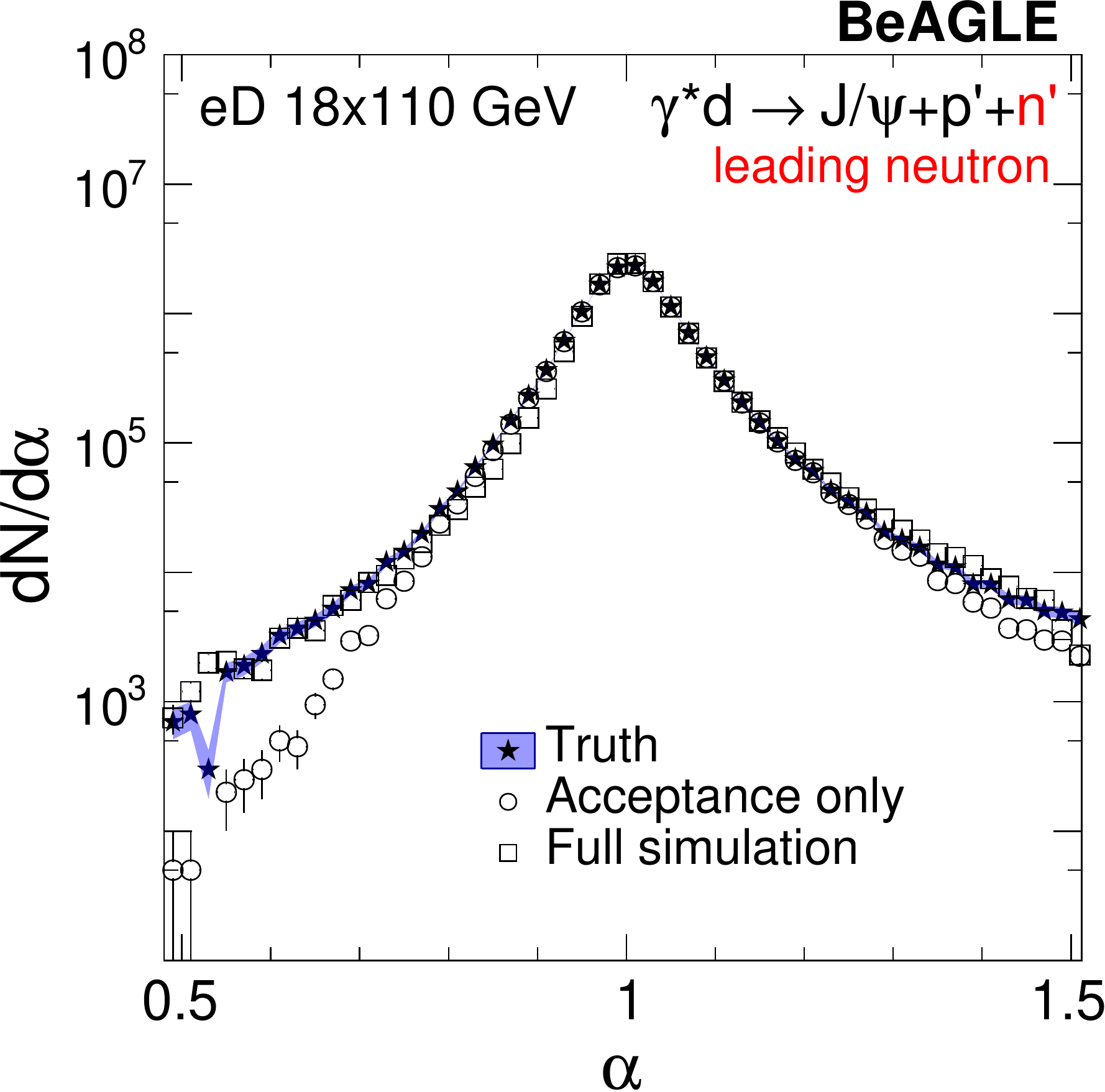}
  \caption{ \label{fig:figure_3} The transverse momentum $p_{\rm{T}}$ (upper row) and the light-cone momentum fraction $\alpha$ (bottom row) of the spectator neutrons (left column) and protons(right column) are shown, respectively, in events associated with incoherent diffractive $J/\psi$ vector meson production in \ed\ collisions. Simulations at the generator level, with acceptances effects only, and for full simulations, are shown with solid, open circles, and open squared markers, respectively.}
\end{figure}

In Fig.~\ref{fig:figure_3}, the transverse momentum $p_{\rm{T}}$ (upper row), and the light-cone momentum fraction $\alpha$ (bottom row), of spectator neutrons (left column) and protons (right column) are shown, for the same set of events as used in Fig.~\ref{fig:figure_2}. The transverse momentum $p_{\rm{T}}$ distributions are found to be robust against the detector resolution even at low $p_{\rm{T}}$ for both cases, as found for the three-momentum $p_{\rm{m}}$ in Fig.~\ref{fig:figure_2}. For the light-cone momentum fraction $\alpha$, similar conclusions can be drawn as in Fig.~\ref{fig:figure_2} by comparing the proton and neutron spectator cases, where the proton spectator tagging has generally the better resolution and the neutron spectator  tagging has better acceptance. The $p_{\rm{T}}$ and $\alpha$ variable of the spectator nucleon parametrize the light-cone spectral function of the deuteron, which will be an important observable for measuring the SRC in deuteron. 

\begin{figure}[thb]
\includegraphics[width=1.65in]{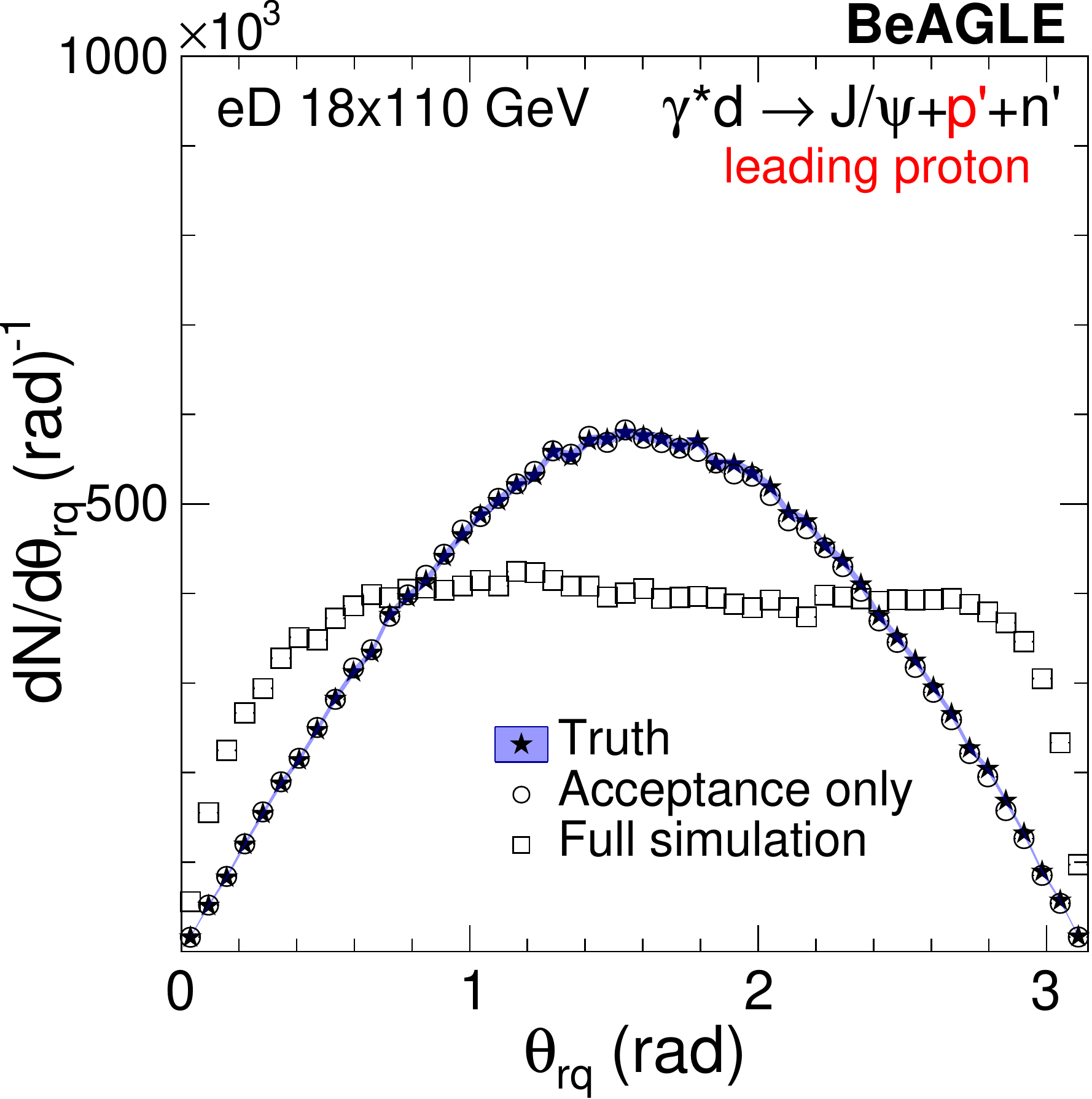}
\includegraphics[width=1.65in]{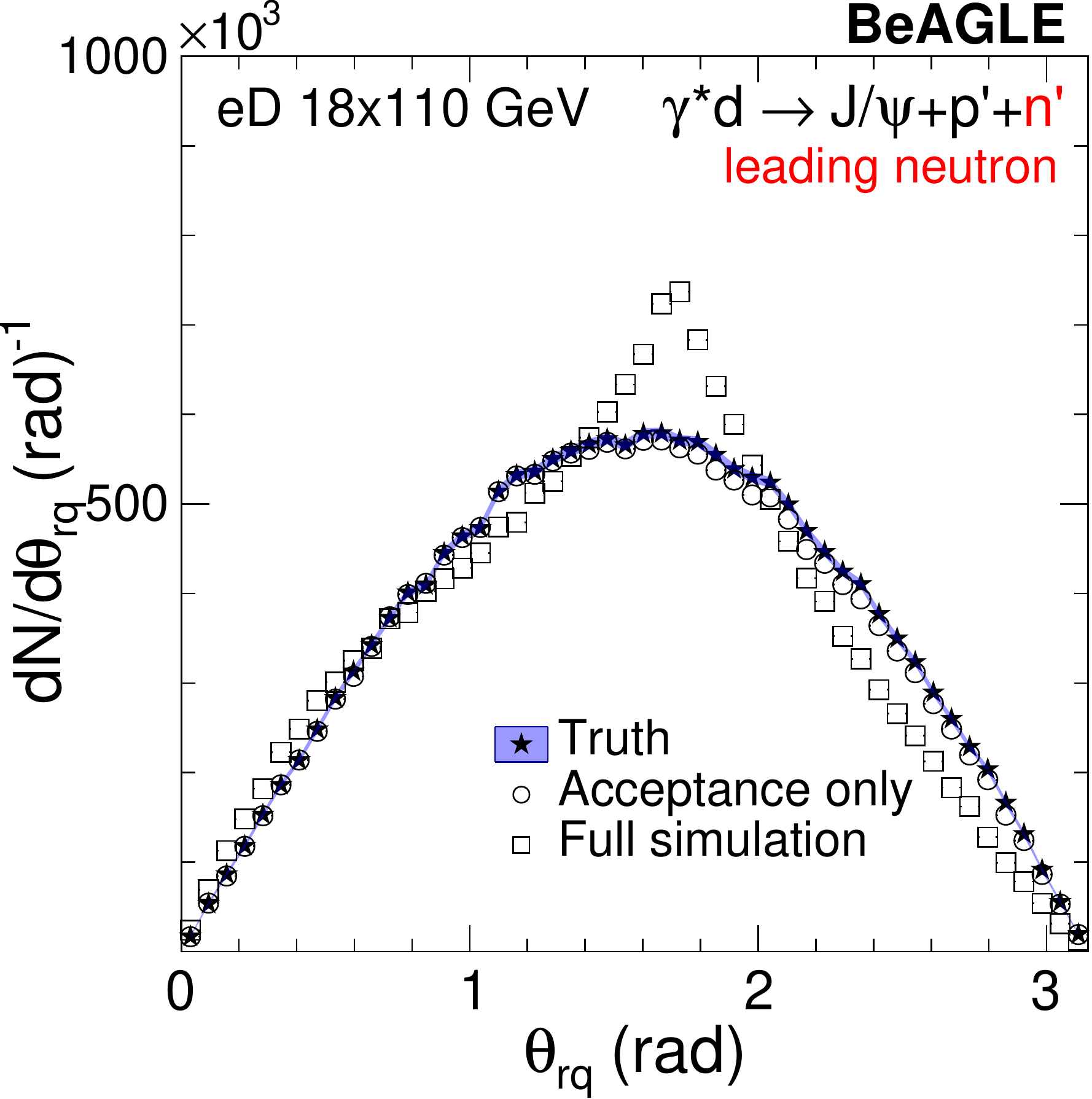}
  \caption{ \label{fig:figure_4} The polar angle between the virtual photon and the spectator neutron (left) and spectator proton (right), respectively, in incoherent diffractive $J/\psi$ vector meson production in \ed\ collisions. These distributions are integrated over all $p_{\rm{m}}$ range. Simulations at the generator level, with acceptance effects only, and with full simulations, are shown with solid, open circles, and open squared markers, respectively.}
\end{figure}

Fig.~\ref{fig:figure_4} shows the polar angle $\theta_{\rm{rq}}$ distributions between the virtual photon and the spectator nucleon in the deuteron rest frame for neutron spectators (left) and proton spectators (right), respectively. These distributions are integrated over the entire $p_{\rm{m}}$ range, which is dominated by the low momentum spectators. Since the events are dominated by low $p_{\rm{m}}$, the acceptance of both cases are found to be similar and close to 100\%. However, after including all effects in the simulation, the $\theta_{\rm{rq}}$ distributions are found to be significantly smeared compared to the generator level ones. This observable has been found in our study to be the most sensitive to detector resolutions.  The differences in the proton and neutron are due to the different smearing effects between the two particles - protons are reconstructed from simulated detector hits; neutrons only have the acceptance applied in the simulation, and a Gaussian smearing applied by hand. The largest component in the $\theta_{\rm{rq}}$ smearing comes from the Lorentz-boost of the deuteron beam momentum, where a small momentum and angular smearing in the lab frame will be enhanced in the deuteron rest frame. For higher $p_{\rm{m}}$ ranges, the difference between the smeared and true distribution becomes smaller, see Sec.~\ref{subsec:misc}.

It is noted that the $\theta_{\rm{rq}}$ distribution is a standard observable for SRC measurements at fixed target experiments~\cite{Egiyan:2007qj}, while the resolution of the $\theta_{\rm{rq}}$ angle generally is much better without the Lorentz-boost to the target rest frame. At the EIC, the Lorentz-boost to the target rest frame is necessary to access the polar angle $\theta_{\rm{rq}}$, which will be an experimental challenge for the measurement at low $p_{\rm{m}}$. Fortunately, probing the high momentum tail of SRC pairs be less problematic at the EIC.

In addition, the $\theta_{\rm{rq}}$ distribution has been found to be sensitive to the FSI for electron-deuteron scattering in fixed target experiments~\cite{Frankfurt:1996xx,Sargsian:2001ax,laget05,Egiyan:2007qj,Sargsian:2009hf,Boeglin:2011mt,Cruz-Torres:2020uke}. The data show that the FSI contribution peaks for $\theta_{\rm{rq}}$ $\sim 70^o$, while it becomes very small at forward ($\sim$0 degree) and backward ($\sim$180 degree) angles. Therefore at the EIC,  $\theta_{\rm{rq}}$ will provide an important handle to suppress the FSI effect when tagging the high momentum spectator nucleon.

\begin{figure}[thb]
\includegraphics[width=1.65in]{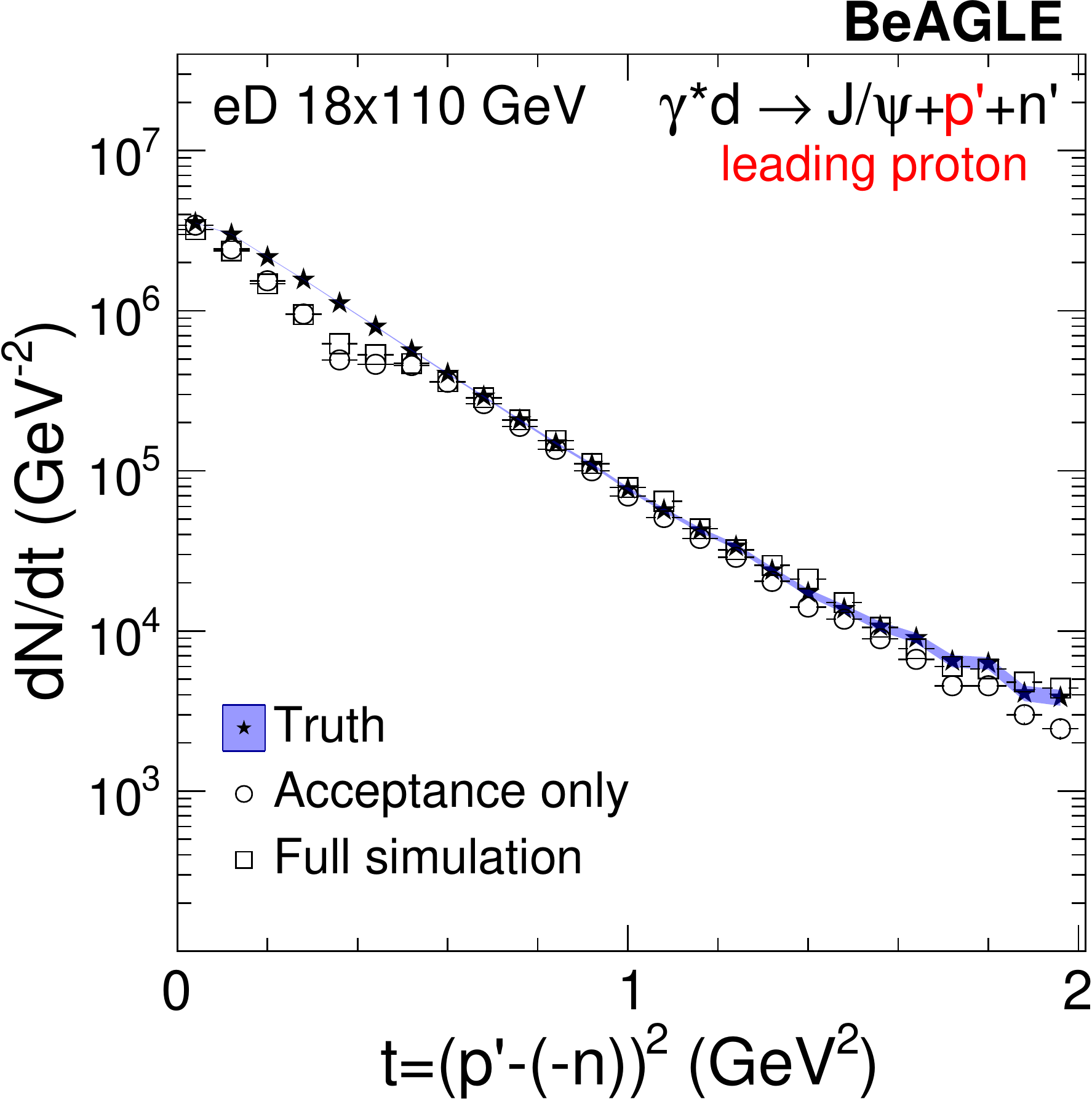}
\includegraphics[width=1.65in]{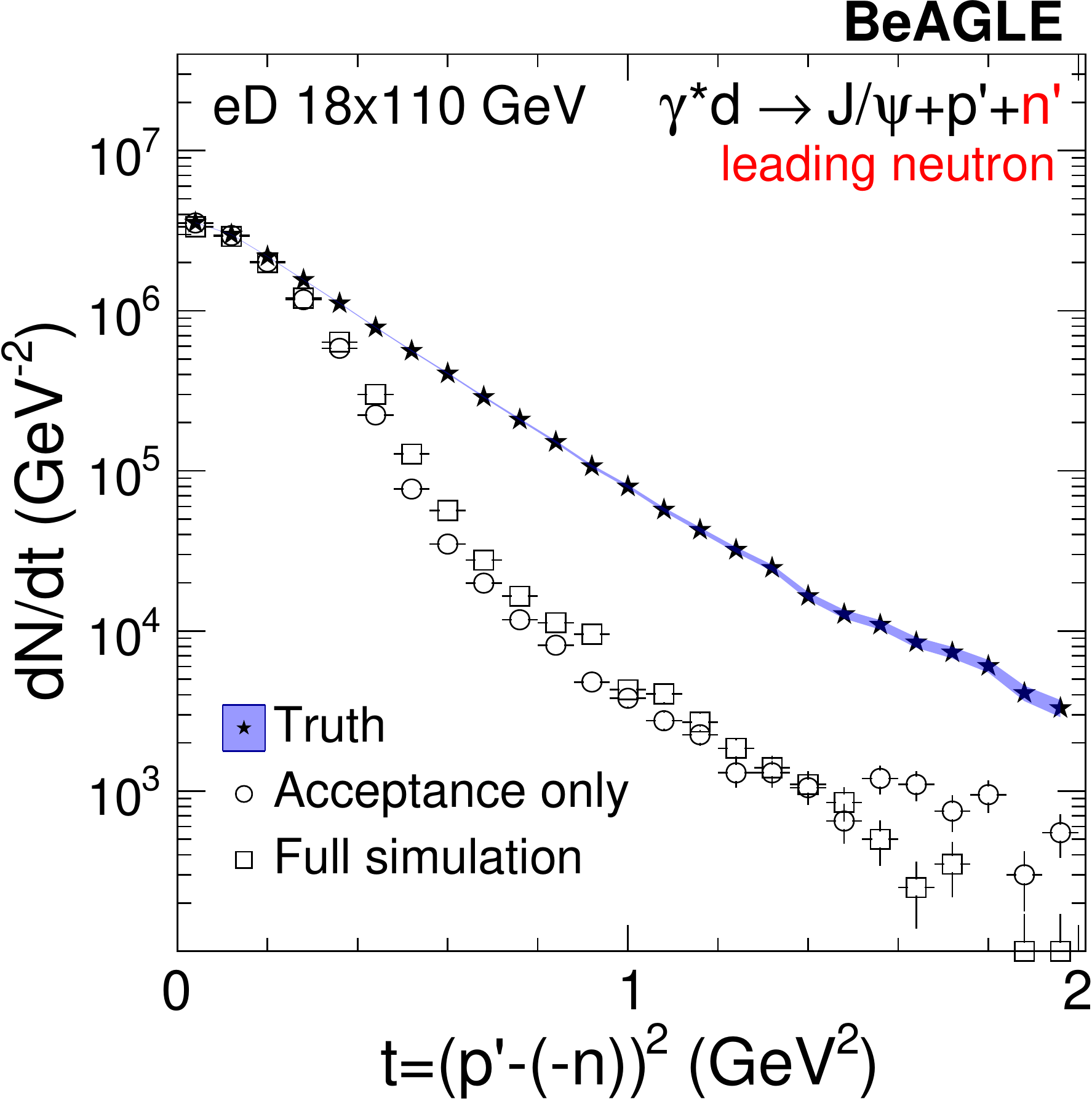}
  \caption{ \label{fig:figure_5} The distribution of the momentum transfer $t$ for neutron spectator (left) and proton spectator (right), respectively, in events associated with incoherent diffractive $J/\psi$ vector meson production in \ed\ collisions. These distributions are integrated over the full $p_{\rm{m}}$ range. Simulations at the generator level, with acceptance effects only, and with full simulations, are shown by solid, open circles, and open squared markers, respectively.}
\end{figure}

Fig.~\ref{fig:figure_5} shows the distribution of the momentum transfer $t$ in incoherent diffractive $J/\psi$ meson production off bound nucleons for the neutron spectator (left) and proton spectator (right) cases respectively. Similar to results shown earlier, the generator level distributions are compared with results including only acceptance effects and for full simulations. Unlike other observables shown earlier, this observable requires double tagging of both leading and spectator nucleons to reconstruct $t$ in the newly proposed method. 

Based on the acceptance-only results, the double-tagging method to reconstruct the $t$ distribution is generally not as good as other observables such as $p_{\rm{m}}$, since it requires good acceptance for both the leading and spectator nucleon. Additionally, the resolution for $t$ is affected by smearing from both reconstructed nucleons. Despite these caveats, the $t$ distribution can be measured up to very high $t$ with good precision utilizing the proposed EIC far-forward detectors, especially in the neutron spectator case. 

Reconstructing the $t$ distributions through method 3 as introduced in Sec.~\ref{sec:kinematics} is found to be generally better than other methods, including the new method proposed in this paper. However, the new method might provide a complementary way of reconstructing the $t$ distributions which is expected to be more robust against QED backgrounds. For comparison among different methods, see the Appendix in Sec.~\ref{sec:appendix} for details.

The generator level $t$ distribution in BeAGLE, reconstructed by the new method using both nucleons, is found to be different from the input $t$ distribution of the PYTHIA-6 $ep$ simulation. Neither the $t$ distributions in PYTHIA 6 nor in the current version of BeAGLE are kinematically correct. The PYTHIA-6 model has no internal nucleon momenta of the incoming nucleon. In BeAGLE, the internal momenta of the bound nucleons, the outgoing nucleon after the interaction are not modified to account for internal nucleon momenta at the initial state in terms of cross sections and kinematic variables.  However a precise  implementation of the momentum transfer $t$ distribution in BeAGLE is not the primary focus of this paper, and it is also not essential for  the conclusions drawn from this study.

At the EIC, the precise measurement of the $t$ distribution in this process is directly related to the gluon imaging of the bound nucleons, similar to the case of general imaging measurements at the EIC. See the EIC white paper for a summary~\cite{Accardi:2012qut}.  Since the $t$ distribution is a Fourier transformation of the source distribution, the measurement of diffractive $J/\psi$ mesons is sensitive to the gluon distribution in the impact parameter space. In this measurement of incoherent diffractive $J/\psi$ meson production with spectator tagging, the gluon distributions of the bound nucleon can be measured for different internal momentum ranges, which might provide a link to the role of gluons in short-range correlations and the modification of gluon structure functions. 

\begin{figure}[thb]
\includegraphics[width=0.49\linewidth]{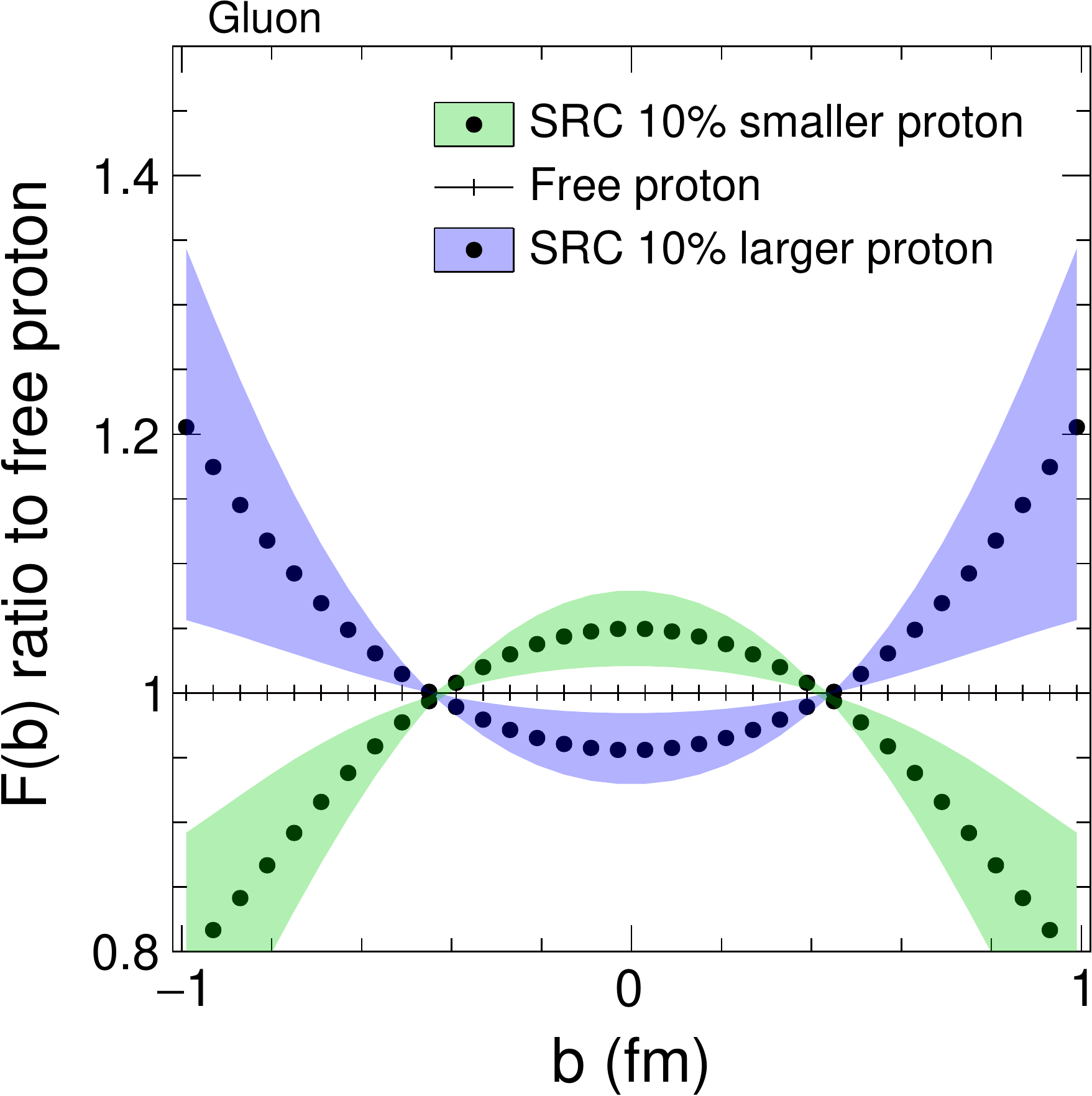}
\includegraphics[width=0.49\linewidth]{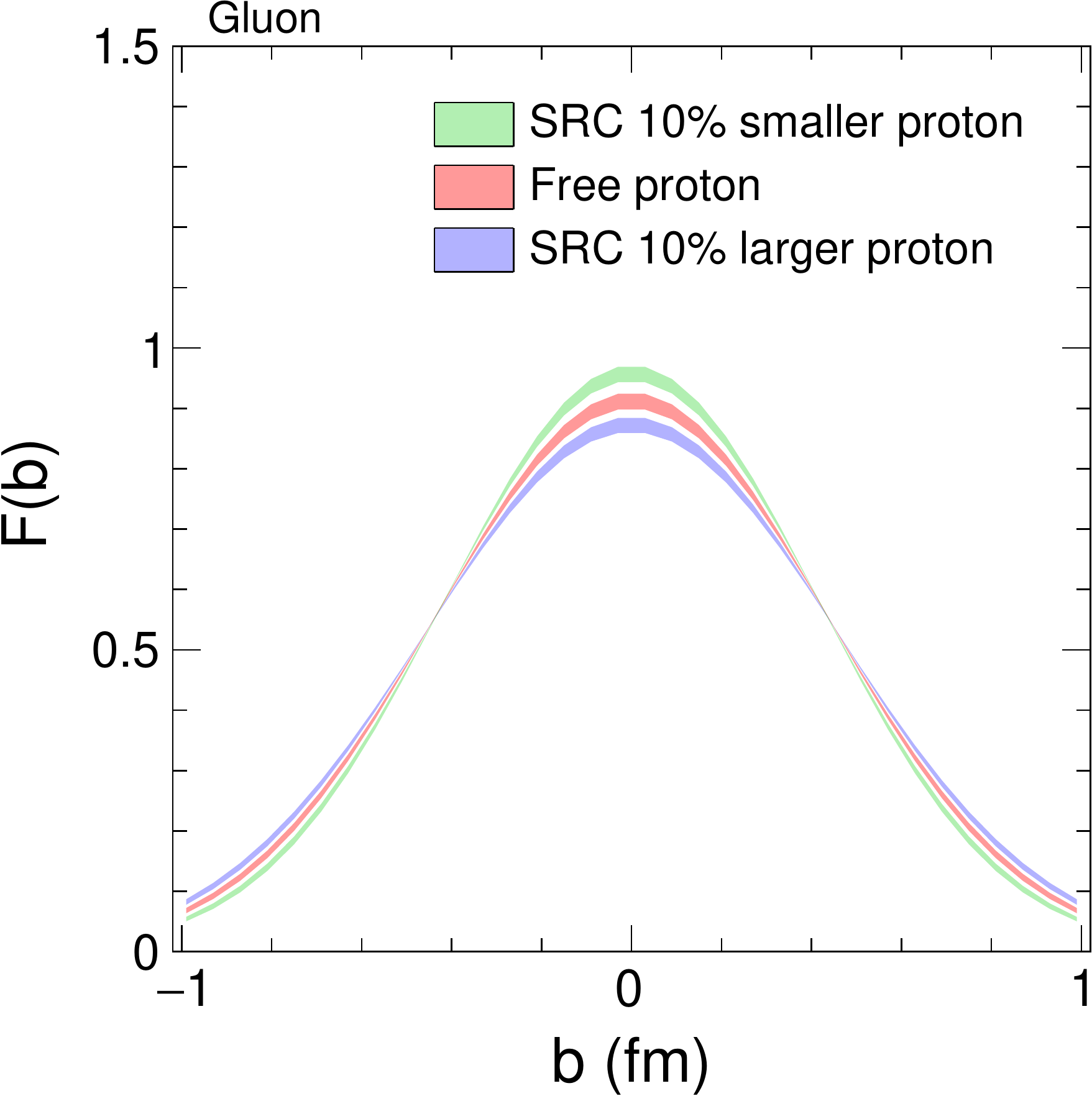}
  \caption{ \label{fig:figure_6} The gluon source distributions $F(b)$ (right) and their ratio between SRC protons and the free proton as a function of impact parameter $b$ (left), based on a Fourier transformation of the $t$ distributions of elastic $J/\psi$ production in $\gamma p$ collisions.  
  The color band indicates a 1$\sigma$ statistical uncertainty.  }
\end{figure}

Even without a fundamental QCD calculation of this possible effect, we
can nevertheless make a qualitative prediction of the SRC-dependent $t$ distribution of the diffractive $J/\psi$ cross section that will reflect the gluon source distributions for different deuteron configurations. Based on Ref.~\cite{Alexa:2013xxa}, the high energy data sample at HERA using the H1 detector  resulted in a measurement of the slope parameter of the  $t$ distribution of $-4.88\pm0.15~\rm{GeV^{-2}}$, with a 3\% uncertainty from the exponential fit. If the deuteron configuration is selected to be in the SRC region, where the bound nucleons are are compressed spatially, this might result in a  gluon source distribution different from the one in a free nucleon. In this paper, we assume a $\pm$ 10\% difference in the nucleon size for events with $p_{\rm{m}}>0.6~\rm{GeV/c}$, shown in Fig.~\ref{fig:figure_6}. In the calculations of Ref.~\cite{Miller:2019mae}, the bound proton in the nucleus is found 
to be larger than a free proton.

With the assumption of a similar statistical precision as obtained by the H1 result~\cite{Alexa:2013xxa}, 
the 10\% difference in the slope parameter of $t$ will result in a 3$\sigma$ significant  different source distribution. This difference will be mostly dominated by the statistical uncertainty, while the systematic uncertainty will be largely, if not fully, canceled. Overall, the significance of the results depends on the signal strength and the statistical uncertainty. For a quantitative prediction, rigorous theoretical calculations are needed. 

In order to achieve a similar statistical precision for $p_{\rm{m}}>0.6~\rm{GeV/c}$ in photoproduction, the integrated luminosity is estimated to be 30 $\rm{fb^{1}}$. For electroproduction of $Q^{2}\sim10~\rm{GeV^{2}}$, 100--500 $\rm{fb^{1}}$ are required depending on the statistical uncertainty of the last measured $t$ bin. For details, see Sec.~\ref{sec:appendix} for luminosity estimations.

Asdie from the gluon imaging of the bound protons with high internal momentum in the deuteron, the spectator tagging technique with deuteron breakup can provide a wide range of applications in three-dimensional imagining of sub-nucleonic structures. For example, deeply virtual Compton scattering (DVCS) on the neutron~\cite{Mazouz:2007aa,Benali:2020vma} can be measured via a similar process in terms of final-state particles as shown in Fig.~\ref{fig:figure_1}, except for a real photon in the final state instead of a $J/\Psi$ particle. The same quantity, $t$ distribution, can be measured in order to access the Generalized Parton Distributions (GPD) of the neutron. The detector requirements that have been studied in this paper will also apply to the analysis  of neutron GPDs. In addition, using the spectator tagging method, the DVCS process on the neutron can be unambiguously identified on an event-by-event basis through the detection of the proton spectator. Note that in Refs.~\cite{Mazouz:2007aa,Benali:2020vma}, the contributions from DVCS on proton and DVCS on deuteron as a whole were statistically subtracted on the cross section level, while the analysis technique proposed in this paper is an active detection of DVCS on the neutron. Finally, related to the short-range correlated pairs of nucleons, the free neutron GPD might be accessible by tagging the proton spectator with on-shell extrapolation, which has never been done before and will be a unique measurement at the EIC. 

\section{\label{sec:conclusion} Conclusions}
Incoherent diffractive $J/\psi$ production with spectator tagging in electron-deuteron collisions at the EIC is investigated using the BeAGLE event generator. Both the leading and spectator nucleons from the deuteron breakup can be detected using the far-forward detectors where the internal nucleon momentum can be reconstructed via the four-momentum of the spectator. Other kinematic variables, such as the momentum transfer $t$ of the leading nucleon, can be reconstructed by tagging both nucleons in the final state.  

Observables that are sensitive to short-range correlations are simulated using the most up-to-date interaction region design and proposed forward detectors at the EIC. Extremely high internal nucleon momentum configurations in the deuteron of  0.8--1 GeV/c can be detected with good acceptance and excellent resolution. The $J/\psi$ photoproduction and electroproduction for different deuteron initial-state configurations can be systematically studied with high statistical precision, given the integrated luminosity anticipated at the EIC. These measurements are expected to be a sensitive probe to the short-distance dynamics arising purely from gluons at the low-$x$ region, and their dependence on internal nucleon configurations. 

\section*{Acknowledgments}
We thank Mark Strikman and Christian Weiss for fruitful discussions on the SRC and deuteron physics. We thank Stefan Schmitt for discussion on the H1 published results. The authors would also like to thank Alexander Kiselev for technical support with EicRoot used for the full simulation studies. The BeAGLE event generator was initiated at Brookhaven National Laboratory~(BNL) with code development and improvements supported by BNL, JLab, and the Laboratory of Nuclear Science at the Massachusetts Institute of Technology (MIT-LNS).
The work of Z.~Tu is supported by LDRD-039 and the Goldhaber Distinguished Fellowship at Brookhaven National Laboratory. 
The Work of Z.~Tu, JH.~Lee, and T.~Ullrich are supported by the U.S. Department of Energy under Award DE-SC0012704. 
The work of A.~Jentsch and E.~Aschenauer is supported by the U.S. Department of Energy under Contract No. de-sc0012704, and A.~Jentsch is also supported by the Program Development program at Brookhaven National Laboratory.
The work of R.~Venugopalan is supported by  the U.S. Department of Energy, Office of Science, Office of Nuclear Physics, under contract No. DE- SC0012704.
The work of M.~Baker was supported by DOE contracts DE-SC0012704, DE-AC05-06OR23177, by MITLNS, and by Jefferson Lab LDRD projects LDRD1706 and LDRD1912.  
The work of L.~Zheng is supported by National Natural Science Foundation of China under Grant No. 11905188.
The work of O.~Hen is supported by DE-SC0020240. 
The work of D.W.~Higinbotham is supported by U.S. Department of Energy contract DE-AC05-06OR23177. 

\bibliography{beagle}

\section{\label{sec:appendix} Appendix}
In this section, there are four subsections to discuss a few technical aspects in detail.In Sec.~\ref{subsec:detector_requirements}, the full breakdown of effects from different detectors and beam-related effects are discussed, along with potential issues that may arise. In Sec.~\ref{subsec:t_reco}, different methods of the $t$ reconstruction are compared, and pros and cons of each method are discussed. In Sec.~\ref{subsec:lumi}, luminosity estimates for this measurement are performed based on the HERA published $ep$ results and the EIC detector concept.

\subsection{\label{subsec:detector_requirements} Detector and beam-related effect simulations}

In the case of the silicon detectors for proton detection, the beam angular divergence is the dominant source of smearing, and is non-correctable in the reconstruction. The beam angular divergence gives a small transverse momentum component to particles within each bunch, where the particles are intended to have a purely longitudinal component. This effect is generally beam-energy dependent (less divergence at lower energy beams), and also dependent on the size of the beam at the collision point. In general, the goal is to reduce the beam size at the collision point to increase luminosity at the expense of an increased beam divergence. 

An additional source of smearing comes from the rotation of the bunches by the crab cavities, which are intended to ensure the bunches collide ``head-on", despite the crossing angle of 25 mrad. This crab cavity rotation has the effect of smearing the size of the collision source by $\sim$(bunch length)*(1/2 crossing angle). This effect is sub-dominant compared to the angular divergence and detector reconstruction effects, and is correctable with precise timing resolution in forward detectors.

The final source of smearing comes from the detector reconstruction. For the protons, the finite size of the silicon pixels introduce smearing in the determination of the coordinate of a ``hit" on the sensor plane. Additionally, the assumption of linear optics was assumed for the transfer of protons from the IR to the off-momentum detectors. This assumption works fine for protons that are not severely off-momentum with respect to the deuteron beam. To limit the effect of this additional smearing, resolutions were calculated only for protons within $\sim$5\% of the peak of the proton momentum distribution. For neutrons, the intrinsic energy and angular resolution of the detector introduces smearing in measured energy and polar angle of the neutrons. 

In this study, it was found that the angular divergence is the largest source of smearing for the protons, while the intrinsic detector smearing is dominant for the neutrons. In both cases, these two effects are the largest, with the beam-energy spread and crab cavity induced vertex smearing being sub-dominant.

\begin{table}
\begin{center}
  \begin{tabular}{|l|c|c|c|c|}
    \hline
    \multirow{2}{*}{$p_{T}$ Resolution} &
      \multicolumn{2}{c|}{Proton} &
      \multicolumn{2}{c|}{Neutron} \\\cline{2-5}
    & \% & MeV/$c$ & \% & MeV/$c$ \\
    \hline
    $p_{T} < 140$ MeV/$c$ & 15 & 22 & 29 & 37 \\
    \hline
    $140 < p_{T} < 350$ MeV/$c$ & 8 & 25 & 14 & 43 \\
    \hline
    $350 < p_{T} < 630$ MeV/$c$ & 6 & 30 & 10 & 52\\
    \hline
    $p_{T} > 630$ MeV/$c$ & 4 & 26 & 9 & 70\\
    \hline
  \end{tabular}
  
  \begin{tabular}{|l|c|c|c|c|}
    \hline
    \multirow{2}{*}{E Resolution} &
      \multicolumn{2}{c|}{Neutron} \\\cline{2-3}
    & \% & GeV/$c$ \\
    \hline
    $50 < p < 80$ GeV/$c$  & 7.5 & 5.5 \\
    \hline
    $80 < p < 110$ GeV/$c$  & 7 & 7 \\
    \hline
    $110 < p < 130$ GeV/$c$  & 6.7 & 8.5 \\
    \hline
    $p > 130$ GeV/$c$  & 6.2 & 11 \\
    \hline
  \end{tabular}
  \end{center}
  \caption{\label{tab:table3} Summary of the total measured resolutions, in percentile ranges and average absolute smearing (in MeV/c and GeV/c), for four different ranges of $p_{T}$ for protons and neutrons, and four ranges of $p$ for neutrons. The $p$-range for the protons was limited to $105 < p < 115$ to reduce artificial increases in smearing due to the current assumption of a linear transfer matrix in the  reconstruction of proton momenta at the off-momentum detectors.  }
\end{table}

\begin{table*}[thb]
\fontsize{8}{14}\selectfont
\begin{center}
\begin{tabular}{|l|c|c|c|c|c|c|c|c|c|}
\hline
\multirow{2}{*}{\textbf{Methods}} &
	\multicolumn{8}{c|}{Momentum transfer $t$ ($\rm GeV^{2}$)} \\\cline{2-9} & 0--0.05 & 0.05--0.11  & 0.11--0.17 & 0.17--0.25 & 0.25--0.35 & 0.35--0.49 & 0.49--0.69 & 0.69--1.20  \\
 \hline
 1. $\delta t/t$ (\%) with $t=(p'-p)^{2}$ & - & - & - & - & - & - & - & -   \\
 \hline
 2. $\delta t/t$ (\%) with $t=(e-e'-V)^{2}$ & $>100$  & $>100$  & $>100$  & $>100$ & $>100$ &$>100$ &$>100$ & $>100$   \\
 \hline
 3. $\delta t/t$ (\%) with $t \approx (p_{\rm{_{T,V}}} + p_{\rm{_{T,e'}}})^{2}$ & 20.3  & 7.8   & 5.8  & 4.8 & 3.9 & 3.4 & 3.0 & 2.5 \\
 \hline
 4. $\delta t/t$ (\%) with $t=(p'-(-n))^{2}$ & 49.6 & 41.6 & 36.2 & 31.6  & 28.2 & 24.4 & 17.9 & 16.0   \\
\hline
 \end{tabular}
 \end{center}
  \caption{\label{tab:table4} Summary of the resolution in momentum transfer $t$ reconstructed for different methods for incoherent diffractive $J/\psi$ meson production in \ed\ collisions.}
 \end{table*}

\subsection{\label{subsec:t_reco} Momentum transfer $t$ reconstruction comparisons}

In Table.~\ref{tab:table4}, different methods of reconstruction of the momentum transfer $t$  are summarized in terms of their resolution as a function of $t$. For the method 2., the resolution at the highest $t$ bin is more than 100\%, so it's not usable by any means. For the method 3.. the resolution is slightly better in photoproduction than in electroproduction, and it has the best resolution among all methods.

\subsection{\label{subsec:lumi} Luminosity Requirements at the EIC}

In this analysis, the $Q^{2}$ dependence of the incoherent diffractive $J/\psi$ meson production is not explicitly studied, since the deuteron dissociation and the kinematics of the spectators are expected to be independent of the event $Q^{2}$. However, at the EIC, this process in terms of the momentum transfer $t$ distribution should be measured deferentially in $Q^{2}$ and $\theta_{\rm{rq}}$ intervals, which will be able to eliminate models with or without a $Q^{2}$ and $\theta_{\rm{rq}}$ dependence. Because of the small cross-section of $J/\psi$ meson production and the high nucleon momentum tail of the deuteron wave function, this analysis will require a high luminosity data sample at the EIC in order to achieve a high-precision measurement. 

The H1 and ZEUS experiment at HERA have measured the elastic $J/\psi$ meson production in both photoproduction and electroproduction~\cite{Chekanov:2009ab,Alexa:2013xxa,Aktas:2005xu}. The statistical precision we are planning to achieve at the very high momentum tail (high $p_{\rm{m}}$) for different $Q^{2}$ and $\theta_{\rm{rq}}$ bins, is to reach the same statistical precision as those HERA results. The statistical precision is directly proportional to the raw $J/\psi$ yield the experiment observes. Therefore, the raw $J/\psi$ yield at the EIC for \ed\ incoherent diffractive scattering is needed, which is related to the following factors:
\begin{itemize}
	\item Detector efficiency; assumed to be at least the same as H1 and ZEUS;
	\item Acceptance in terms of $W_{\gamma^{*} N}$ range, directly depends on the photon energy, hadron beam energy, and the $J/\psi$ mass;
	\item Total photon flux in the $W_{\gamma^{*} N}$ range and $Q^{2}$ dependence;
	\item Elastic $J/\psi$ cross section in $\gamma^{*} p$ system; assume the same for $\gamma^{*} n$.
\end{itemize} 

The photon energy in the lab frame in the elastic $J/\psi$ production in $ep$ collisions is defined as, 
\be
k=\frac{1}{2}M_{J/\psi}e^{-y_{_{\rm{J}}}}. 
\ee
\noindent where the $y_{_{\rm{J}}}$ and $M_{\rm J/\psi}$ is the rapidity and mass of $J/\psi$ meson.
The center-of-mass energy $W_{\gamma^{*}p}$ is then, 
\be
W^{2}_{\gamma^{*}p} = 4E_{\rm p}k,
\ee
\noindent where $E_{\rm p}$ is the proton beam energy.
For the incoherent diffractive $J/\psi$ meson production in \ed\ scattering, the same kinematics can be used for calculating the rapidity distribution of $J/\psi$ and the $W_{\gamma^{*}N}$ by knowing the nucleon beam momentum of the deuteron beam. Note that the internal momentum of the nucleon can be different from the beam momentum, the resulting $W_{\gamma^{*}N}$ will be larger or smaller depending on the incoming nucleon momentum, while the rapidity distribution mostly depends on the photon energy. 

In this paper, we are using a conservative selection on the inelasticity $y$ range within $0.01<y<0.85$ for estimating the photon energy range at the EIC. The comparison between the EIC and the published H1 results~\cite{Alexa:2013xxa} by assuming an integrated luminosity of 130~$\rm{pb^{-1}}$ is summarized in Table.~\ref{tab:table5}. Note that in a different publication as in Ref.~\cite{Aktas:2005xu}, the H1 Collaboration reported a wider range of $W_{\gamma^{*}p}$ utilizing a  backward calorimeter to reconstruct the decay products. 

Based on the raw $J/\psi$ yield at the EIC, the projected number of events for photoproduction and one selected bin of $Q^{2}$, $8<Q^{2}<12.7~\rm{GeV^{2}}$, in electroproduction are calculated, in bins of $p_{\rm m}$ and $\theta_{\rm{rq}}$. The $Q^{2}$ dependence of the elastic $J/\psi$ cross section is based on the H1 result from Ref.~\cite{Aktas:2005xu}, where the total photon flux is calculated based on Refs~\cite{Sjostrand:2006za}  assuming $Q^{2}/W^{2}$ is small. The main conclusions on the luminosity requirements are the following, based on the same statistical precision as the published H1 result~\cite{Alexa:2013xxa} for the $t$ distributions.
\begin{itemize}
\item  Photoproduction: spectator nucleon momentum $p_{\rm{m}}$ between 600--800 $\rm{MeV/c}$, 30 $\rm fb^{-1}$ is needed for integrated $\theta_{\rm{rq}}$, and more than 50 $\rm fb^{-1}$ for $\theta_{\rm{rq}}>2 ~\rm{rad}$. 
\item Electroproduction with $8<Q^{2}<12.7~\rm{GeV^{2}}$: spectator nucleon momentum $p_{\rm{m}}$ between 600--800 $\rm{MeV/c}$, 500 $\rm fb^{-1}$ is needed for integrated $\theta_{\rm{rq}}$, and more than 1000 $\rm fb^{-1}$ for $\theta_{\rm{rq}}>2 ~\rm{rad}$. 
\item If the statistical precision for the highest $t$ bin (0.69--1.2 $\rm GeV^{2}$) is aimed at 30\% instead of 13\% as in Ref.~\cite{Alexa:2013xxa}, for electroproduction with $8<Q^{2}<12.7~\rm{GeV^{2}}$, the luminosity requirement becomes 100 and 200 $\rm fb^{-1}$ for integrated $\theta_{\rm{rq}}$ and $\theta_{\rm{rq}}>2 ~\rm{rad}$, respectively.
\end{itemize}
Overall, this particular process requires a high luminosity data sample at the EIC. 

\begin{table*}[thb]
\fontsize{11}{14}\selectfont
\begin{center}
\begin{tabular}{|l|c|c|}
\hline
\multirow{2}{*}{ Comparisons } &
	\multicolumn{2}{c|}{ Integrated luminosity $130~\rm{pb^{-1}}$} \\\cline{2-3} & H1 experiment at HERA & EIC detector ($-4<\eta<4$) \\
 \hline
Electron energy in [GeV]  & 27.6 &  18     \\
 \hline
Photon energy range in [GeV]  & [0.43,3.24] &  [0.18,15.3]     \\
\hline
Proton (neutron) energy in [GeV]  & 920 &  110/nucleon     \\
 \hline
$W_{\gamma^{*}N}$ range in [GeV] & [40,110] & [9,82]       \\
 \hline
$J/\psi$ rapidity range  & [-0.75,1.27]  & [-2.3,2.2]   \\
 \hline
Transversely polarized photon flux  & 0.094 & 0.174        \\
\hline
$\sigma_{\gamma^{*}p}$ of elastic $J/\psi$ integrated over $W_{\gamma^{*}N}$ in [nb] & 4976 & 3634 \\
\hline
Total raw $J/\psi$ yield (both $ee$ and $\mu\mu$ channels)  & 53593$\pm$231 &  74000$\pm$270    \\
\hline
 \end{tabular}
 \end{center}
  \caption{\label{tab:table5} Summary of relevant factors for estimating raw $J/\psi$ yield at the EIC based on the H1 experiment}
 \end{table*}

\subsection{\label{subsec:misc} Other configurations of EIC detectors and energies}

\begin{figure}[thb]
\includegraphics[width=\linewidth]{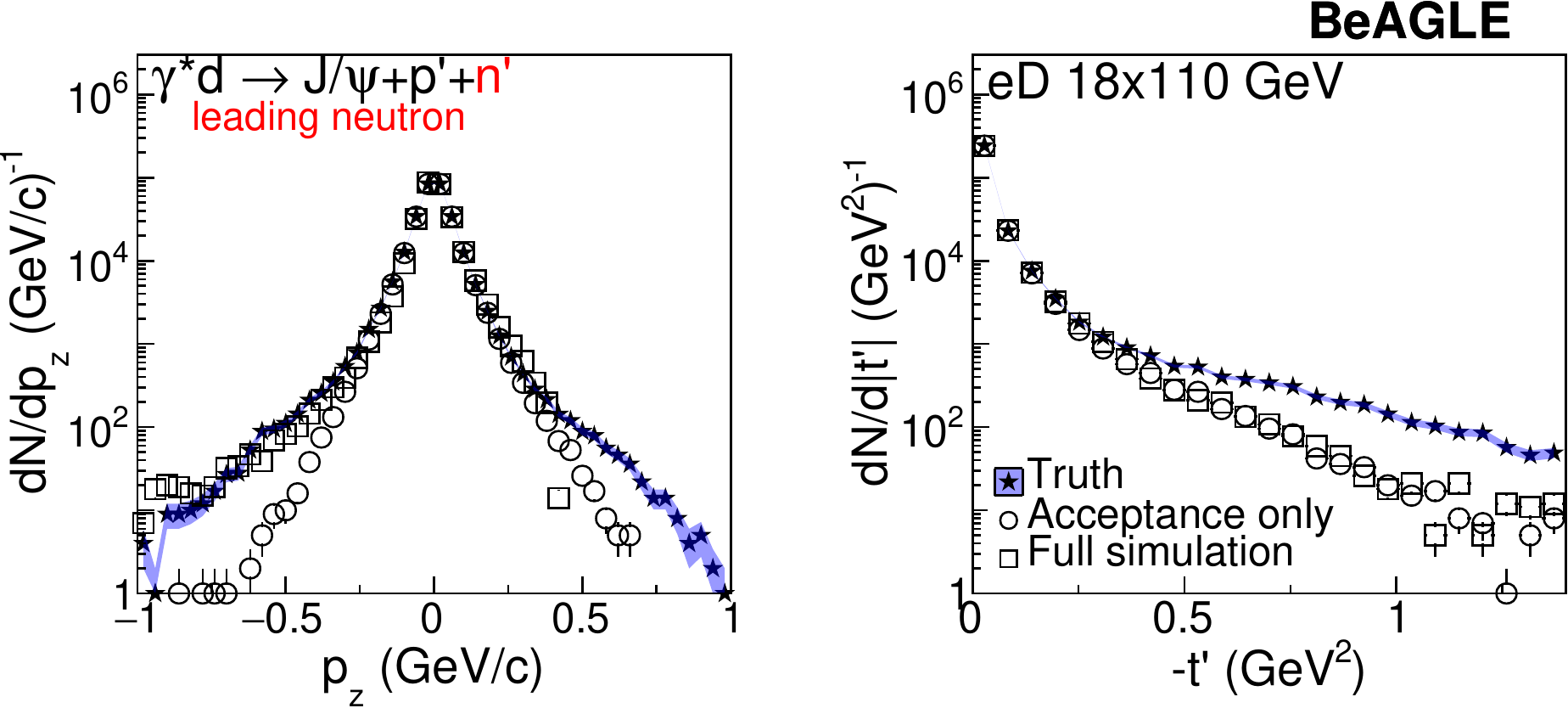}
  \caption{ \label{fig:moreKinePlots_proton} The z-component of the momentum in the ion rest frame (left column) and the off-shellness, t' (right column) for the spectator protons.}
\end{figure}

\begin{figure}[htb]
\includegraphics[width=\linewidth]{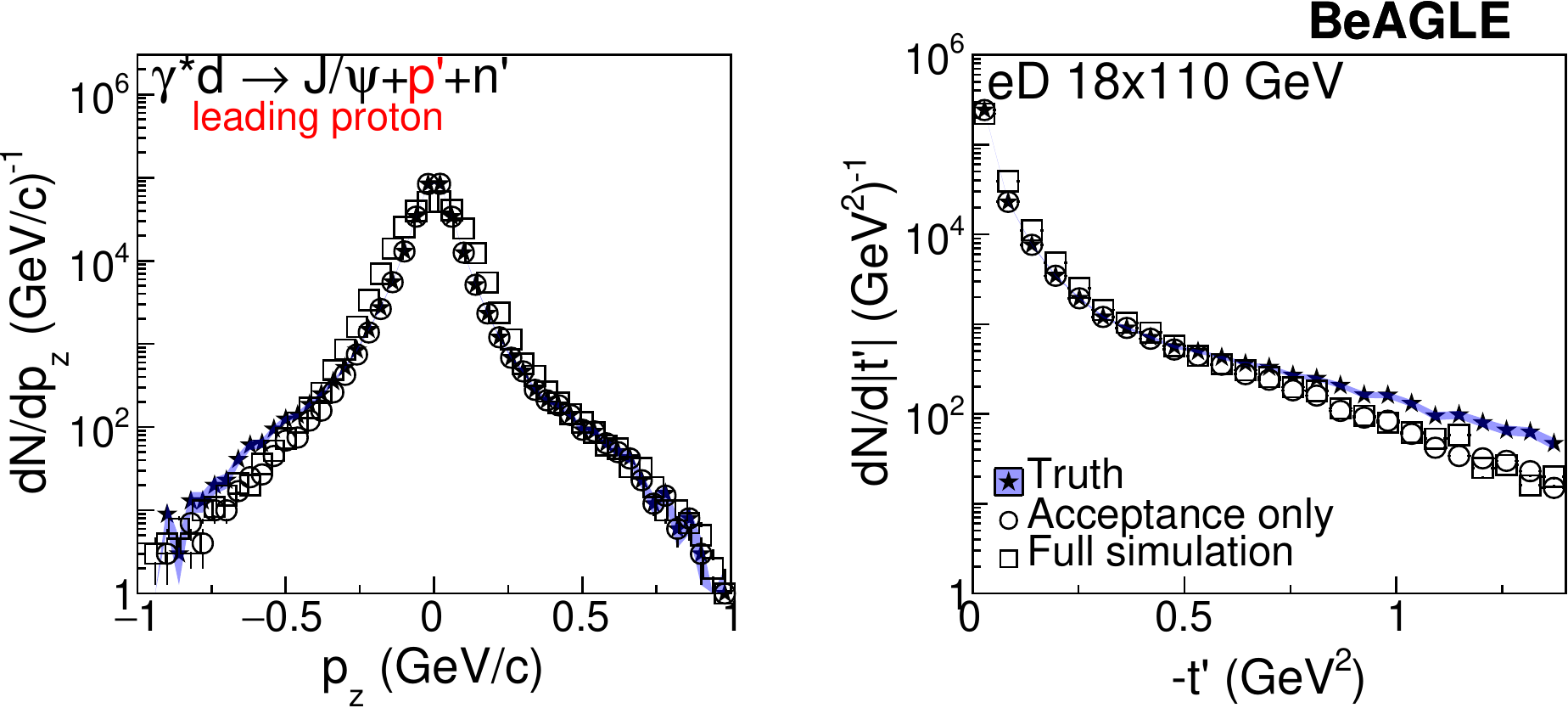}
\caption{ \label{fig:moreKinePlots_neutron} The z-component of the momentum in the ion rest frame (left column) and the off-shellness, t' (right column) for the spectator neutrons.}
\end{figure}

In this subsection, results are presented based on a few other simulation configurations, which would provide more insight for future analysis. In Figs. \ref{fig:moreKinePlots_proton} and \ref{fig:moreKinePlots_neutron}, the distributions of the z-component of the spectator nucleon in the ion rest frame, as well as the off-shellness of the nucleon (t') are shown for 18 GeV electron scatters off 110 GeV deuteron.

\begin{figure}[htb]
\includegraphics[width=0.49\linewidth]{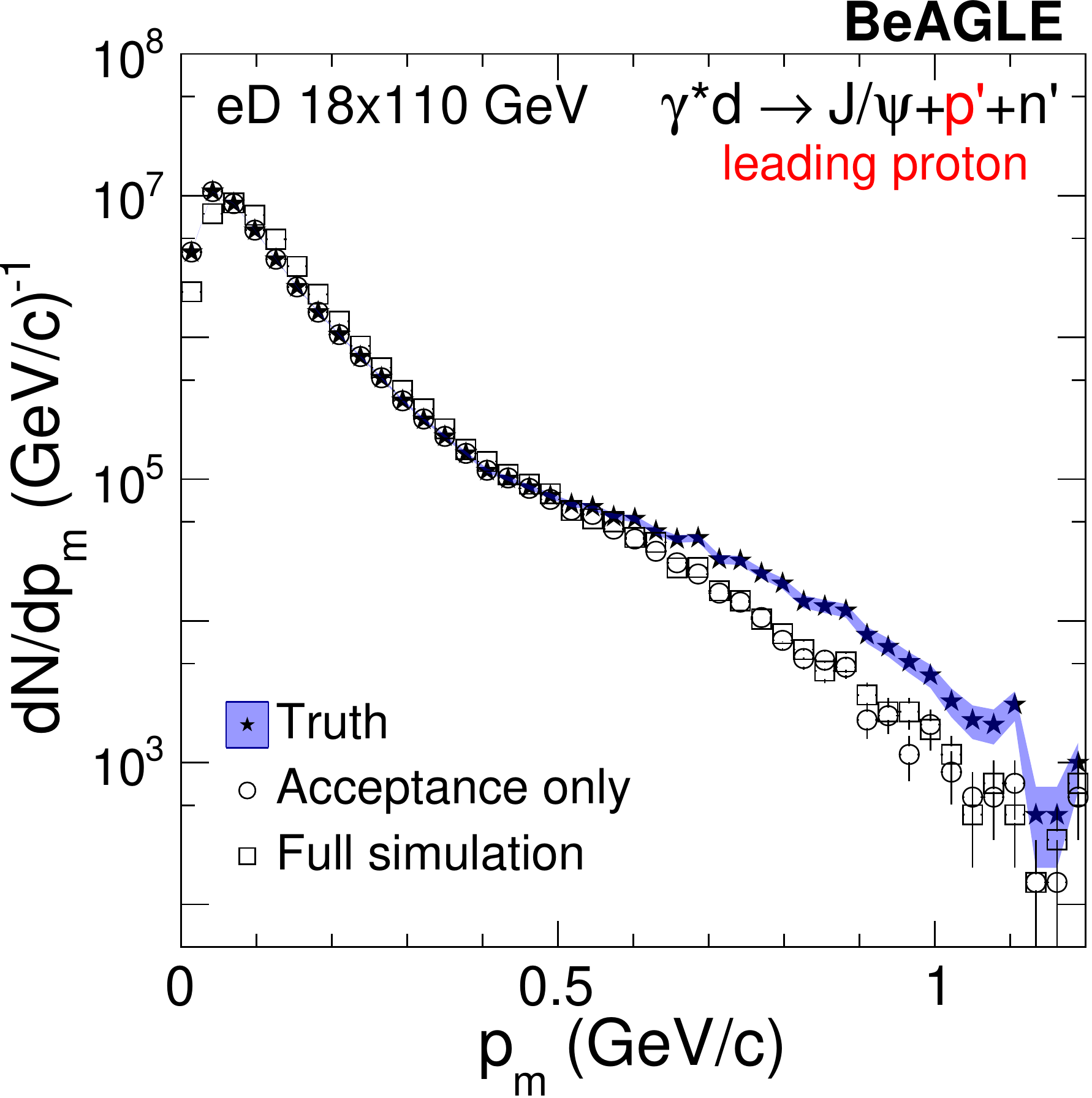}
\includegraphics[width=0.49\linewidth]{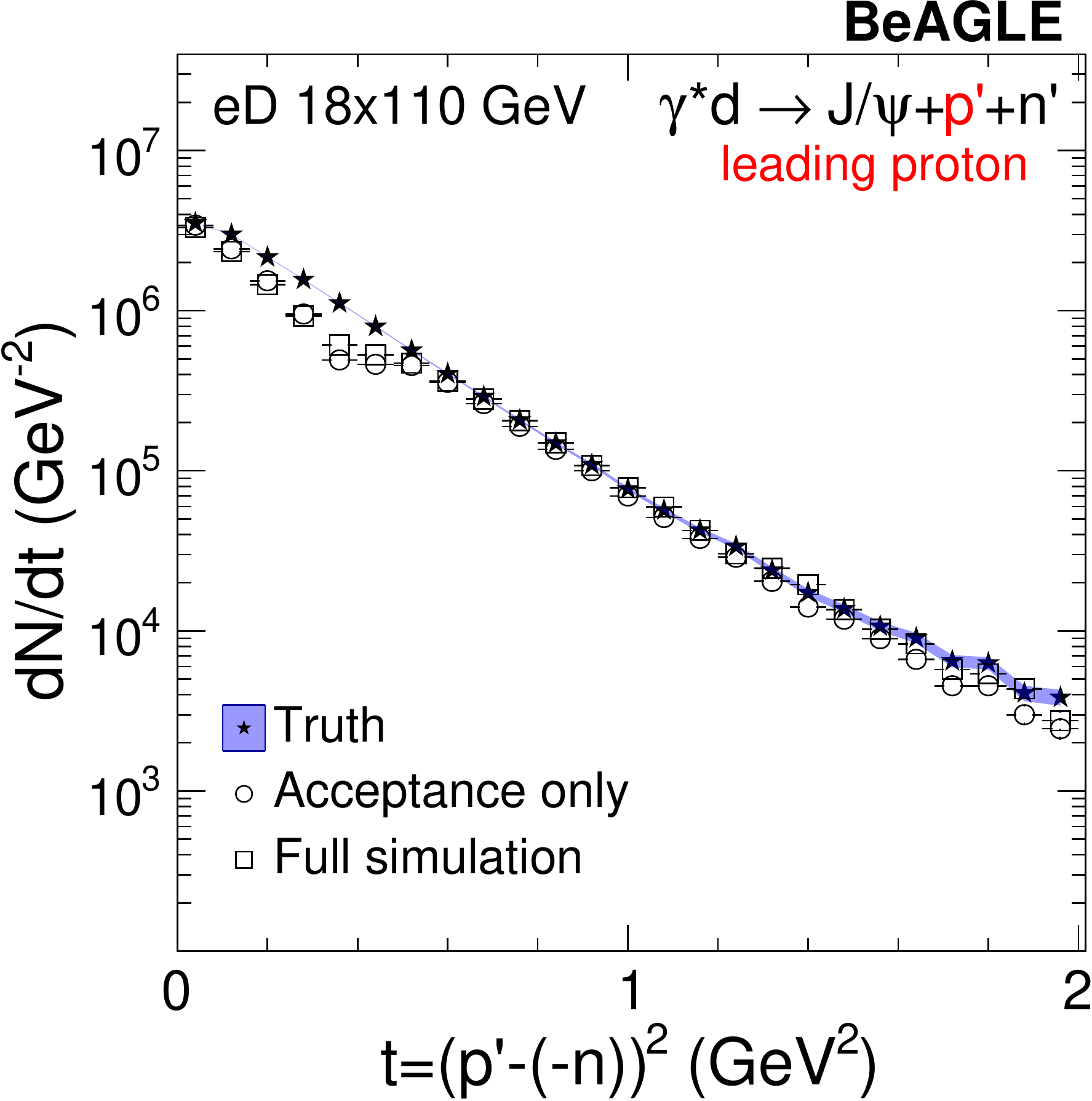}
  \caption{ \label{fig:betterZDC} Plots of $p_{m}$ (left panel) and momentum transfer, t (right panel) for the leading proton case using a ZDC with improved energy resolution ($\frac{\sigma_{E}}{E} = \frac{35\%}{\sqrt{E}} + 2\%$).}
\end{figure}

In the main simulation analysis presented in Sec. \ref{sec:result}, the ZDC energy resolution was chosen to be $\frac{\sigma_{E}}{E} = \frac{50\%}{\sqrt{E}} + 5\%$. Fig. \ref{fig:betterZDC} shows the $p_{m}$ and t-distribution for the leading proton case, but with a ZDC assuming improved energy resolution of $\frac{\sigma_{E}}{E} = \frac{35\%}{\sqrt{E}} + 2\%$. With this improved ZDC resolution, less distortion of the low-momentum portion of the $p_{m}$-distribution, and of the t-distributions are observed compared to the nominal case.

\begin{figure}[htb]
\includegraphics[width=\linewidth]{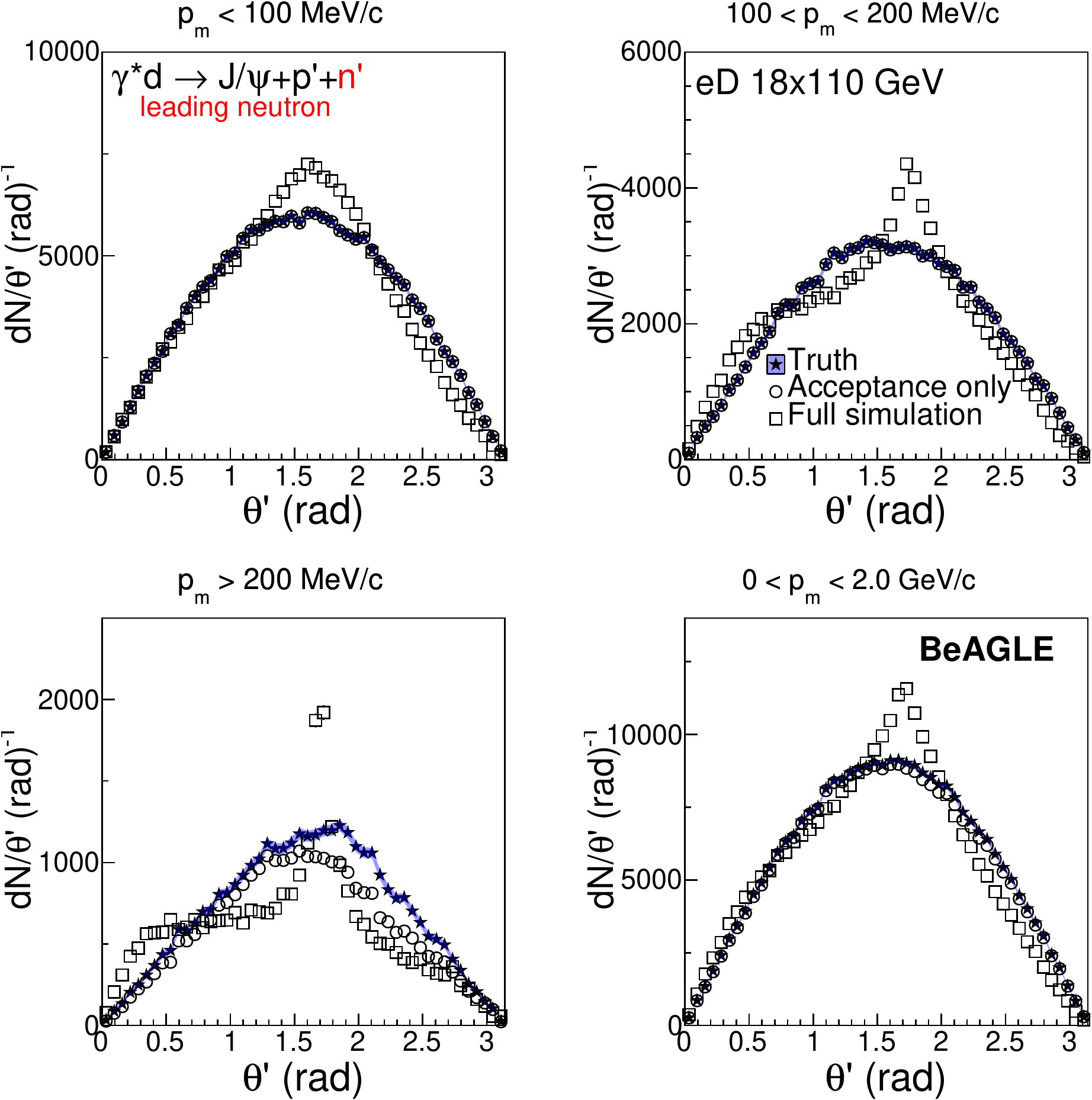}
  \caption{ \label{fig:thetaPrime_vs_pm_proton} $\theta '$-distribution for four ranges of $p_{m}$ (top-left: $p_{m} < 100$ MeV/$c$, top-right: $ 100 < p_{m} < 200$ MeV/$c$, bottom-left: $p_{m} > 200$ MeV/$c$, bottom-right: $0 < p_{m} < 2$ GeV/$c$) for the proton spectator.}
\end{figure}

\begin{figure}[htb]
\includegraphics[width=\linewidth]{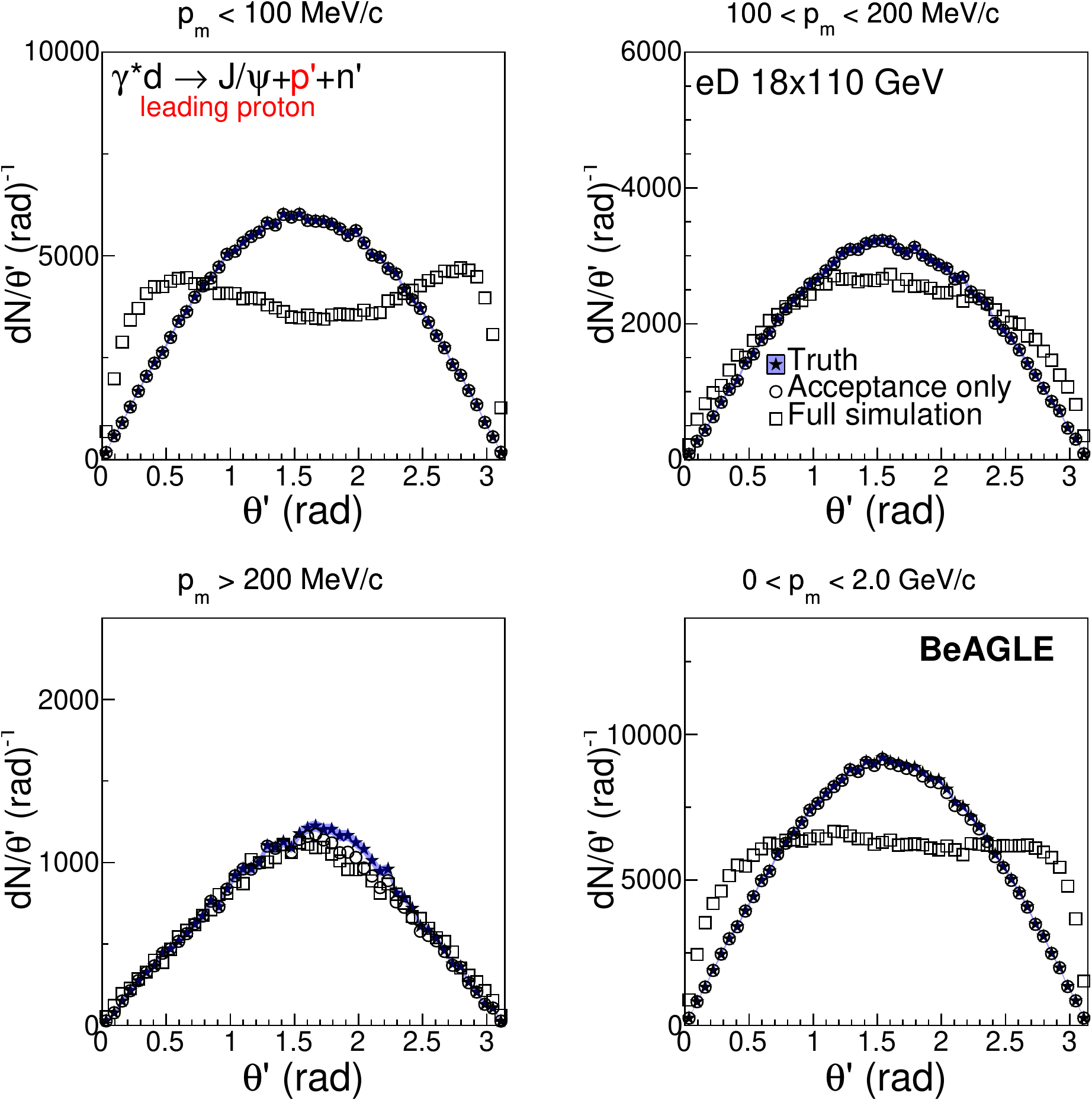}
\caption{ \label{fig:thetaPrime_vs_pm_neutron} $\theta '$-distribution for four ranges of $p_{m}$ (top-left: $p_{m} < 100$ MeV/$c$, top-right: $ 100 < p_{m} < 200$ MeV/$c$, bottom-left: $p_{m} > 200$ MeV/$c$, bottom-right: $0 < p_{m} < 2$ GeV/$c$) for the neutron spectator.}
\end{figure}

Fig. \ref{fig:figure_4} depicts the polar angle between the virtual photon and the spectator nucleon ($\theta '$), however this distribution is sensitive to the choice of $p_{m}$ range, and could be potentially measured more precisely in a specified range of $p_{m}$. Figs. \ref{fig:thetaPrime_vs_pm_proton} and \ref{fig:thetaPrime_vs_pm_neutron} show the $\theta '$-distribution in four ranges of $p_{m}$ for both nucleon spectator cases. The failure different change of shape of the proton spectator $\theta '$-distribution compared to the neutron case is due to the assumption of linear optics for the transfer matrix used to reconstruct severely off-momentum protons at the off-momentum detectors. This distortion becomes worse at higher values of $p_{m}$.

\begin{figure}[htb]
\includegraphics[width=0.49\linewidth]{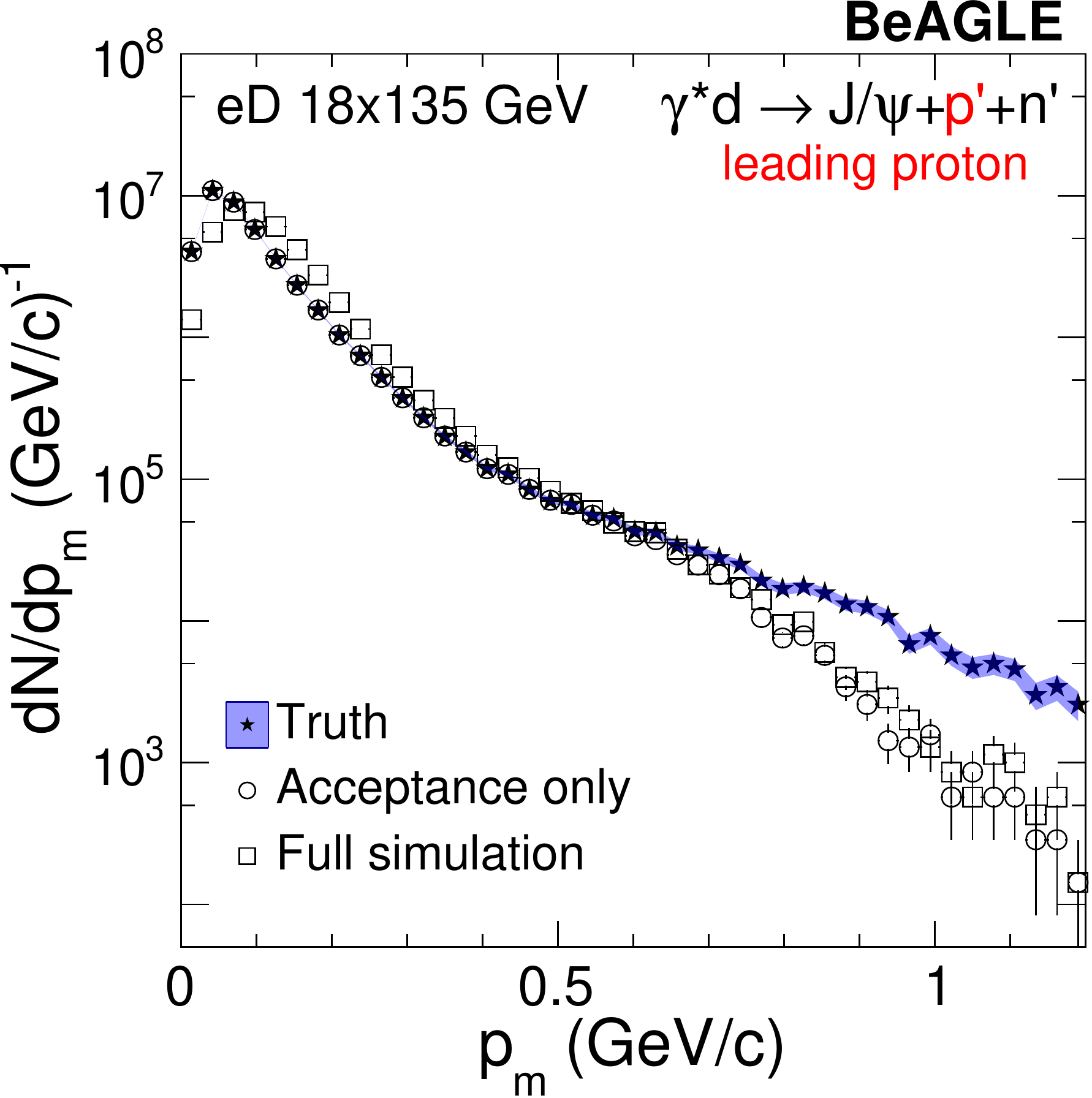}
\includegraphics[width=0.49\linewidth]{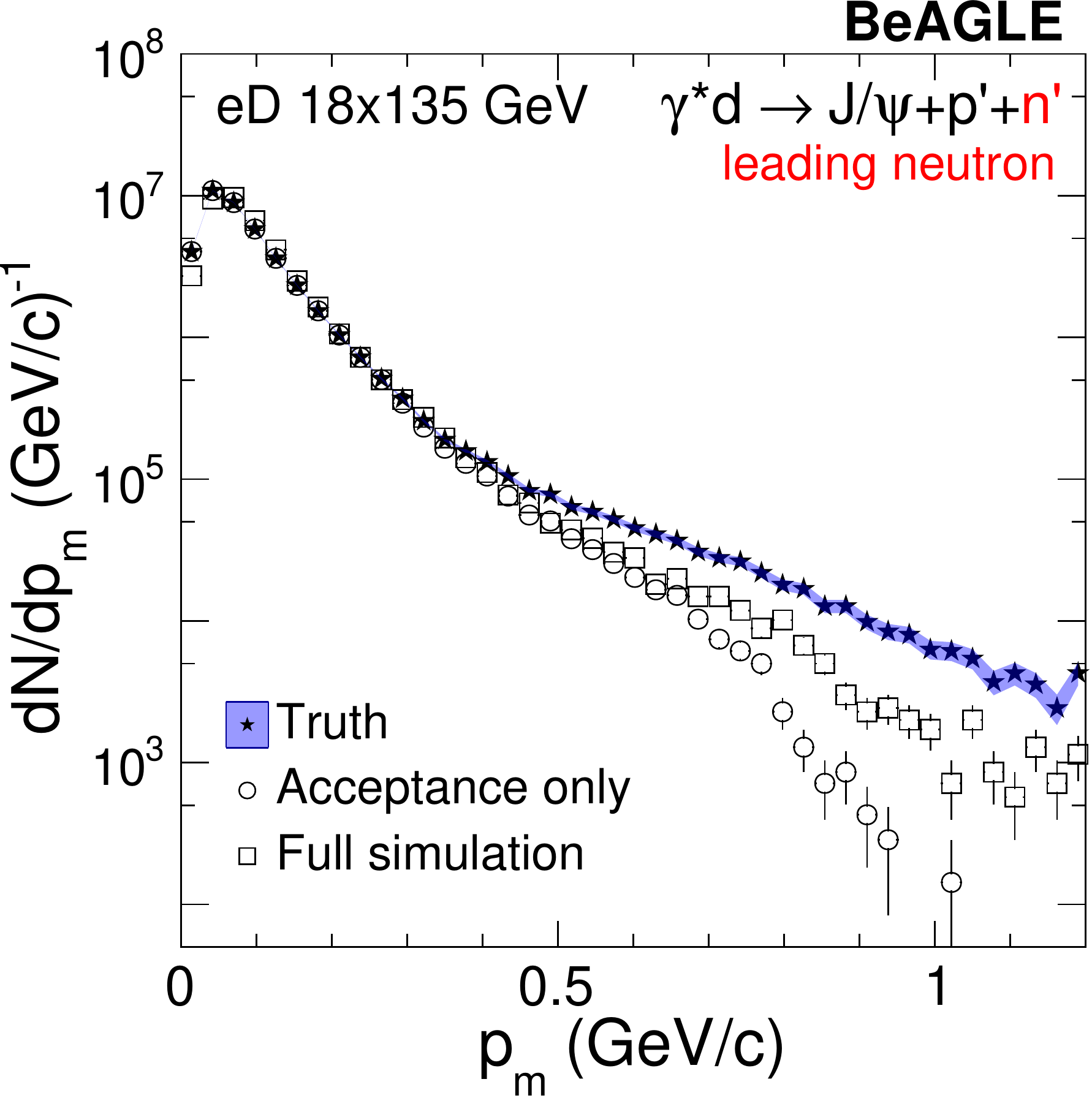}
  \caption{ \label{fig:pm135GeVPlots} Plots of $p_{m}$ for the leading proton case (left) and leading neutron case (right) at energy configuration of 18 GeV electron scatters off 135 GeV deuteron, are presented.}
\end{figure}

\begin{figure}[htb]
\includegraphics[width=0.49\linewidth]{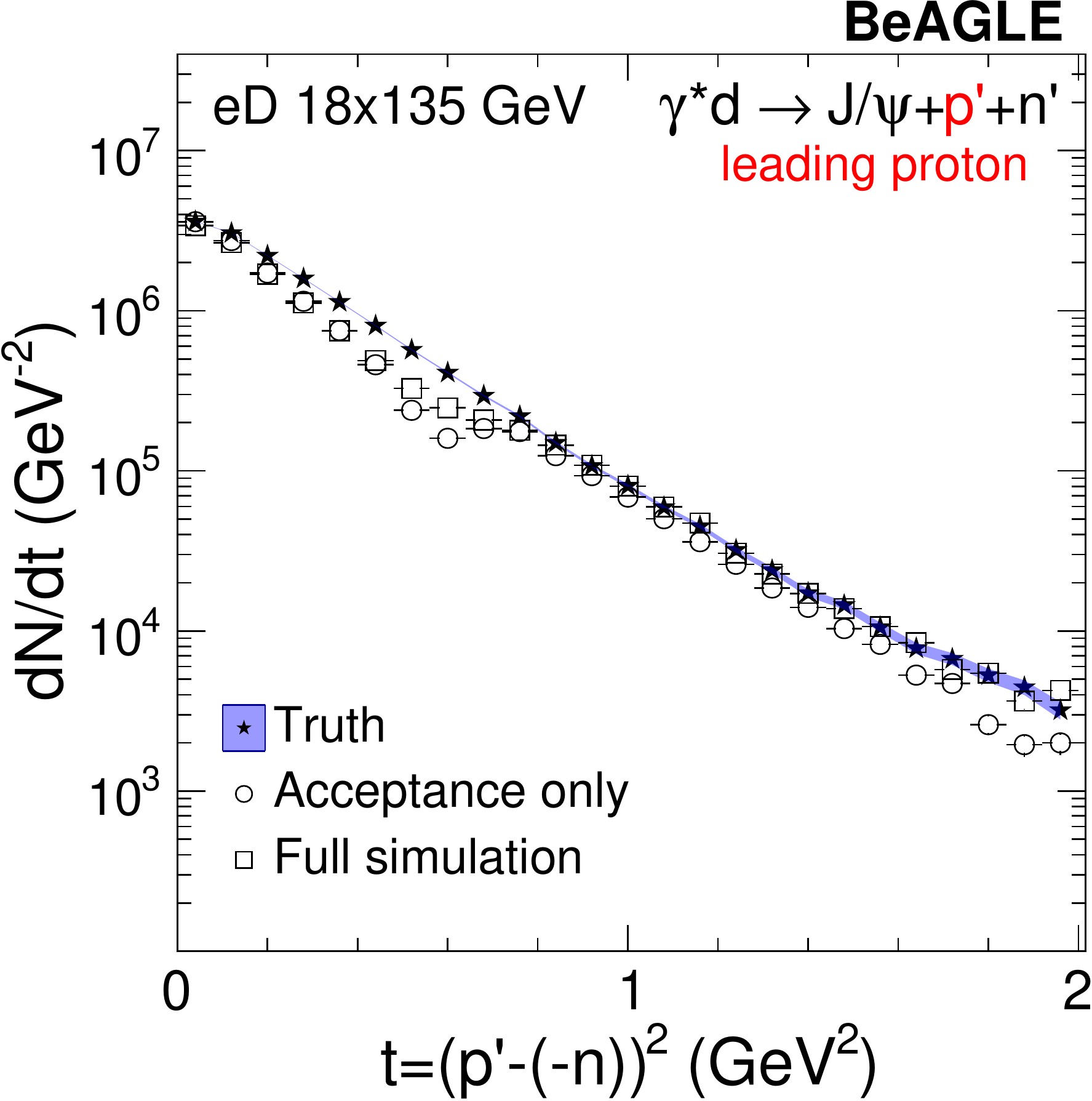}
\includegraphics[width=0.49\linewidth]{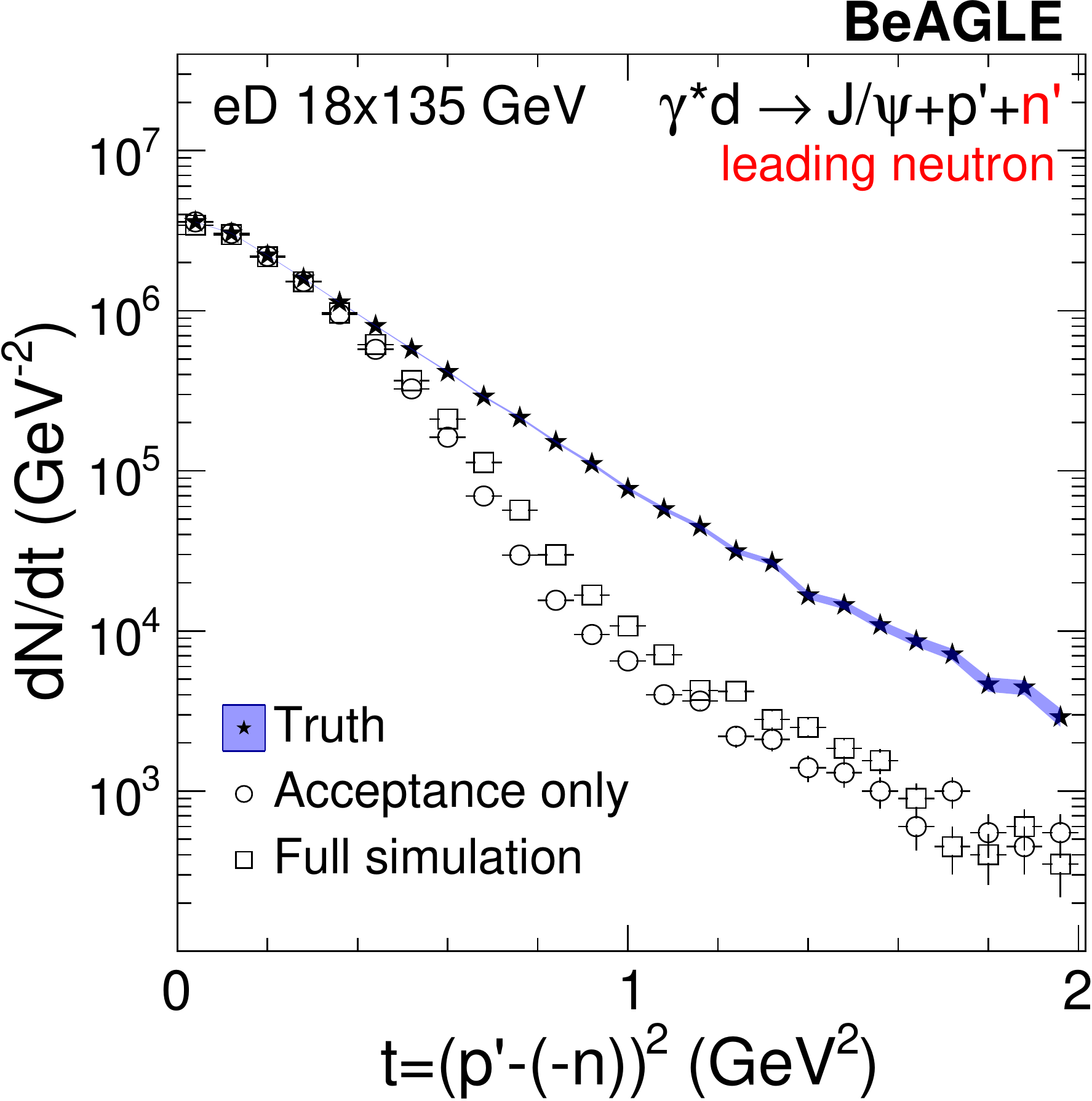}
  \caption{ \label{fig:t135GeVPlots} Plots of $t$ distributions for the leading proton case (left) and leading neutron case (right) at energy configuration of 18 GeV electron scatters off 135 GeV deuteron, are presented.}
\end{figure}

Finally, since the EIC design allows for deuteron beam energies up to 135 GeV/n, simulations of e+d 18x135 GeV/n were also performed. The physics observable, e.g., the $p_{m}$ and $t$ distributions are shown in Fig. \ref{fig:pm135GeVPlots} and in Fig. \ref{fig:t135GeVPlots}, respectively. Further studies will be carried out at lower energies in the future (namely, e+d 5x41 GeV/n), where the proton acceptance begins to be dominated by larger scattering angles, an therefore further employing the B0 detector, while the neutron acceptance is reduced due to larger-$theta$ neutrons being lost in the magnet aperture before reaching the ZDC. 

\end{document}